\documentclass[11pt,a4paper]{article}
\pdfoutput=1
\usepackage{jcappub}
\usepackage{bm}
\usepackage{indentfirst}
\usepackage{amsmath}
\usepackage{graphicx}
\usepackage{float}
\usepackage{amssymb}
\usepackage{subfigure}
\usepackage{amssymb}
\usepackage{psfrag}
\usepackage{hyperref}
\usepackage{tensor}

\usepackage{color}
\usepackage{comment}
\usepackage{CJK}
\usepackage{array}
\usepackage{indentfirst}

\begin{document}
\title{Gravitational waves from eccentric extreme mass-ratio inspirals as probes of scalar fields}
\author[a]{Chao Zhang,}
\author[b,1]{Yungui Gong,\note{Corresponding author.}}
\author[c]{Dicong Liang,}
\author[d,e]{and Bin Wang}
\affiliation[a]{School of Physics and Astronomy, Shanghai Jiao Tong University, 800 Dongchuan Rd, Shanghai 200240, China}
\affiliation[b]{School of Physics, Huazhong University of Science and Technology, 1037 LuoYu Rd, Wuhan, Hubei 430074, China}
\affiliation[c]{Kavli Institute for Astronomy and Astrophysics, Peking University, No.5 Yiheyuan Road , Beijing
100871, China}
\affiliation[d]{Center for Gravitation and Cosmology, Yangzhou University, 88 South Daxue Rd, Yangzhou, Jiangsu 225009, China}
\affiliation[e]{School of Aeronautics and Astronautics, 800 Dongchuan Rd, Shanghai 200240, China}
\emailAdd{zhangchao666@sjtu.edu.cn}
\emailAdd{yggong@hust.edu.cn}
\emailAdd{dcliang@pku.edu.cn}
\emailAdd{wang\_b@sjtu.edu.cn}

\keywords{gravitational waves, EMRIs, scalar charge}
\abstract{
We study eccentric orbits of the Schwarzschild spacetime for extreme mass ratio system (EMRI) in modified gravity theories with additional scalar fields.
Due to the additional energy and angular momentum carried away by the scalar field,
the orbit of the EMRI in modified gravity decays faster than that in general relativity.
The time that it takes the eccentricity $e$ to reach the minimum is shorter and the values of the semi-latus rectum $p$ and $e$ at the turning point when $e$ reaches the minimum are bigger for larger scalar charge $d$.
In addition to the calculation of energy fluxes with numerical methods,
we also use the Post-Newtonian expansion of the rate of energy carried away by the scalar field in eccentric orbits to understand the behaviors of the energy emission.
By adding the scalar flux to the open code FastEMRIWaveforms of the Black Hole Perturbation Toolkit,
we numerically generate fast gravitational waveforms for eccentric EMRIs with scalar fields
and use the faithfulness between waveforms with and without the scalar charge to discuss the detection of scalar charge $d$.
The detection error of the scalar charge is also estimated with the method of the Fisher information matrix.}

\maketitle

\section{Introduction}
In 2015 the Laser Interferometer Gravitational-Wave Observatory (LIGO) Scientific Collaboration and the Virgo Collaboration directly detect the first gravitational wave (GW) event GW150914 \cite{Abbott:2016blz, TheLIGOScientific:2016agk},
this opens a new window to understand the nature of gravity in the nonlinear and strong field regimes.
So far there have been tens of confirmed GW detections  \cite{LIGOScientific:2018mvr, LIGOScientific:2020ibl, LIGOScientific:2021usb, LIGOScientific:2021djp}.
However, the ground-based GW observatories can measure GWs in the frequency range $10-10^3$ Hz only,
due to the seismic and gravity gradient noises,
which means that a wealth of astrophysical signals in the lower frequency ranges $10^{-4}-10^{-1}$ Hz are difficult to be explored.
Extreme mass ratio inspirals (EMRIs) composed of a stellar-mass object inspiraling into a supermassive black hole (SMBH) with masses in the range of $10^6-10^9~M_{\odot}$ \cite{Kormendy:1995er}, emitting GWs in the milli-Hertz band,
are the most promising GW sources for space-based observatories like the Laser Interferometer Space Antenna (LISA) \cite{Danzmann:1997hm, Audley:2017drz},
TianQin \cite{Luo:2015ght} and Taiji \cite{Hu:2017mde,Gong:2021gvw}.
GW signals from EMRIs provide highly accurate information about the parameters and orbits of the system which enable us to constrain additional gravitational fields in alternative theories of gravity \cite{Maselli:2020zgv, Maselli:2021men, Yunes:2011aa, Dai:2021olt, Zhang:2022hbt, Liang:2022gdk, Zhang:2023vok, Guo:2023mhq}.

In the framework of Einstein's general relativity (GR),
GWs have tensor polarizations only and quadrupolar radiation is the lowest mode,
while additional polarizations and emission channels might exist in alternative theories of gravity \cite{eardley1975, Will:1977wq}.
For example, in scalar-tensor theories such as Brans-Dicke (BD) theory \cite{Brans:1961sx, Dicke:1961gz,eardley1975},
there exists the scalar dipolar radiation in quasicircular orbits \cite{Alsing:2011er,Damour:1996ke,Saijo:1996iz,Gerard:2001fm,Alsing:2011er,Yagi:2015oca}. In addition to the dipolar emission, monopolar radiation also presents if the eccentricity is nonzero \cite{Will:1989sk,Brunetti:1998cc},
such additional emission channels in BD theory for eccentric binaries can be helpful to distinguish BD theory from GR.
The detection of eccentric binaries and the measurement of the eccentricity are not only useful to infer the binary formation mechanism \cite{Mroue:2010re,Gondan:2017hbp,Gondan:2018khr,Lower:2018seu,Romero-Shaw:2019itr,Romero-Shaw:2020siz,Lenon:2020oza,Wu:2020zwr,Zevin:2021rtf,Tucker:2021mvo}, but also helpful to constrain
alternative theories of gravity.
Although the strength of the scalar mode is not comparable with those of the tensor modes, there are still cumulative effects that can be detected by the space-based detectors \cite{eardley1975,Will:1977wq,Will:1989sk,Damour:1992we,Damour:1998jk,Alsing:2011er,Antoniadis:2013pzd,Zhang:2019hos,Seymour:2019tir,Maselli:2020zgv,Maselli:2021men,Jiang:2021htl,Guo:2022euk}.
The black hole (BH) endowed with the additional scalar field can  carry a scalar charge whose quantity is inverse to the mass squared in a class of scalar-tensor theories \cite{Campbell:1991kz,Mignemi:1992nt,Kanti:1995vq,Yunes:2011we,Kleihaus:2011tg,Sotiriou:2013qea,Sotiriou:2014pfa,Antoniou:2017acq,Doneva:2017bvd,Silva:2017uqg,Cardoso:2020iji,Nair:2019iur}.
Different theories and models may have different constraints on the scalar charge.
A notable example is (decoupled) Einstein-dilaton Gauss-Bonnet gravity \cite{Yagi:2015oca, Nair:2019iur},
the scalar charge-to-mass ratio $d$ of the BH is proportional to the dimensionless coupling constant of the theory $\beta=2\alpha/\rm{mass}^2$, specifically $d=2\beta+(73/15)\beta^3$ \cite{Julie:2019sab}.
The constant $\alpha$ with a unit of $(\rm length)^2$ describes the coupling between quadratic-in-curvature scalars and a massless scalar field.
For the existence of the stable BH, the charge $d$ should be less than $0.572$ \cite{Pani:2009wy}.
The constraint through the Shapiro time delay using the Saturn probe Cassini on $\sqrt{\alpha}$ is limited below $8.9\times10^6~\rm{km}$ \cite{Bertotti:2003rm}. 
The low-mass x-ray binary GRO J0422+32 \cite{Kanti:1995vq} and A0620-00 \cite{Yagi:2012gp} constrain the $\sqrt{\alpha}$ below $3.1~\rm{km}$ and $1.9~\rm{km}$, respectively.
This new constraint is more than six orders of magnitude stronger than the solar system bound.
The strongest constraint from the Bayesian analysis on GW190814, assuming it is a BBH merger, is $\sqrt{\alpha}\leq 5.6~\rm{km}$ for using GWTC-1  \cite{Nair:2019iur}, $\sqrt{\alpha}\leq 0.4~\rm{km}$ for using GWTC-2  \cite{Wang:2021jfc}
and $\sqrt{\alpha}\leq 0.27~\rm{km}$ for using GWTC-3 \cite{Wang:2023wgv} at $90\%$ credibility.
If GW190814 is not a BH binary, then the strongest constraint comes from GW 200115, giving $\sqrt{\alpha}\leq 1.1~\rm{km}$ \cite{Wang:2023wgv}, which corresponds to $d\leq 0.022$ for the BH with mass $10~M_{\odot}$.

For EMRIs, the radiation reaction acting on the test particle can be split into two parts: the dissipative one and the conservative one.
When the motion of the test particle is geodesic over a time scale comparable to the orbital period,
we can consider the dissipative part under the adiabatic approximation only.
The dissipative part can be calculated from the energy and angular momentum flux at the horizon of the central BH and at infinity.
The method of calculating the effects of radiation reaction on the bound orbits in the Schwarzschild spacetime is based on the Teukolsky formalism for BH perturbations \cite{Teukolsky:1973ha, Press:1973zz, Teukolsky:1974yv, Zerilli:1970se, Chandrasekhar:1975nkd}.
A more detailed presentation can be found in \cite{Poisson:1993vp, Poisson:1995vs, Apostolatos:1993nu,Cutler:1994pb}.
When the adiabatic approximation is valid, the calculation proceeds as follows \cite{Cutler:1994pb}.
Assuming that the motion is strictly geodesic over several orbital periods,
we calculate the energy and angular momentum flux from the system and then give the time-averaged rates of change of the orbital parameters.
Then we derive the orbital evolutions for the semi-latus rectum $p$ and the eccentricity $e$ from the initial conditions.
As pointed in Ref. \cite{Cutler:1994pb},
as long as the extreme mass ratio is suitably satisfied,
the results are compatible with the initial assumption, and the calculation is self-consistent.

Maselli et al. discussed the possible detection of scalar fields with EMRIs onto Schwarzschild BHs in circular orbits by LISA
and they found that the accumulated dephasing due to the scalar charge of the small body with four-year observations of EMRIs can be detected \cite{Maselli:2020zgv}.
By calculating the difference in the number of GW cycles accumulated by EMRIs on circular orbits with the mass $\mu=10M_\odot$ for the small body,
they found that the smaller mass the supermassive BH,
the smaller the scalar charge $d$ detectable \cite{Maselli:2020zgv}.
Then the analysis was extended to EMRIs onto $10^6M_\odot$ Kerr BHs with the dimensionless spin $\chi=0.9$ in circular orbits \cite{Maselli:2021men}.
It was found that with one-year observation, LISA could detect the scalar charge $d$
as small as $d\sim 5\times 10^{-3}$ and $d\gtrsim 0.01$ by the dephasing and
faithfulness between two GWs with and without the scalar charge, respectively \cite{Maselli:2021men}.
Using the Fisher information matrix (FIM) method,
one-year observation of EMRIs with a signal-to-noise ratio (SNR) of 150,
LISA could constrain the scalar charge as small as $d\sim 0.05$ to be inconsistent with zero at the $3\sigma$ confidence level and the scalar charge is highly correlated with the mass of the small compact object and anti-correlated with the spin parameter and the mass of the central Kerr BH \cite{Maselli:2021men}.
Barsanti et al. numerically calculated the evolution of eccentric orbits in Kerr spacetime and they discussed how the scalar charge affects the orbital evolution for different eccentricities and different values of the BH spin,
they also discussed the effects of the scalar charge and the eccentricity on the dephasing and faithfulness between signals with and without the scalar charge,
and they found that the eccentricity improves the detectability of the scalar charge,
at least for $d\gtrsim 0.01$ \cite{Barsanti:2022ana}.
Since more massive Schwarzschild BH in circular EMRIs improves the distinguishability
of the scalar charge while massive Kerr BH decreases the distinguishability
of the scalar charge, can the eccentricity affect the dependence of the detectability of the scalar charge on the mass of the Schwarzschild BH?
In this paper, we address this problem by calculating GWs in eccentric orbits and by using the Fisher information matrix to estimate the errors of detecting the scalar charge.
We also use the post-Newtonian expansion of GW emission by the scalar field
to understand the numerical results and evaluate factors that influence the detectability of the scalar field.
Furthermore, based on the open code FastEMRIWaveforms \cite{Katz:2021yft},
we add the energy flux emitted by scalar fields to generate fast GW waveforms for eccentric EMRIs.
In Sec. \ref{sec2}, we introduce the background of the model and the Teukolsky perturbation formalism.
In Sec. \ref{sec3}, we calculate the source of the Teukolsky equation and orbital evolution equations as well as waveforms.
In Sec. \ref{sec4}, we give the numerical results of orbital evolution and waveform.
In Sec. \ref{sec5}, we give the analytical results of energy rates.
In Sec. \ref{sec6}, we use the modified FastEMRIWaveforms to generate GW waveforms and perform the Fisher information matrix to estimate the errors of detecting scalar charge.
The last section is devoted to conclusions and discussions.

\section{Theoretical framework}
\label{sec2}
In this section, we consider a general action \cite{Maselli:2020zgv}
\begin{equation}
  S[\mathbf{g},\varphi,\Psi]=S_0[\mathbf{g},\varphi]+
 \alpha S_c[\mathbf{g},\varphi]+S_m[\mathbf{g},\varphi,\Psi]\,,\label{action}
\end{equation}
where
\begin{equation}
  \label{S_0}
S_0= \int d^4x \frac{\sqrt{-g}}{16\pi}\left(R-\frac{1}{2}\partial_\mu\varphi\partial^\mu\varphi\right)\,,
\end{equation}
$R$ is the Ricci scalar and $\varphi$ is a massless scalar field.
The nonminimal coupling between the
metric tensor $\mathbf{g}$ and the scalar field $\varphi$ is denoted by the term $\alpha S_c$ and the coupling constant $\alpha$ has the dimension
$[\alpha]=(\text{mass})^n$ with $n>0$.
In the skeletonized approach \cite{1975ApJ...196L..59E, Damour:1992we, Julie:2017ucp, Julie:2017rpw}, the secondary body in an EMRI is treated as a point particle,
the matter action $S_m$ for the secondary body reads
\begin{equation}
  \label{skeleton_action}
  S_m[\mathbf{g},\varphi,\Psi]=-\int m(\varphi)d\tau,
\end{equation}
where $\tau$ denotes the proper time and the mass $m(\varphi)$ of the point particle depends on the value of the scalar field at the location of the particle.
Varying the action with respect to the metric tensor and the scalar field yields
\begin{equation}
  G^{\mu\nu}=R^{\mu\nu}-\frac{1}{2}g^{\mu\nu}R= T_{{\rm scal}}^{\mu\nu} +
  \alpha T_{c}^{\mu\nu}+ 8\pi T_p^{\mu\nu}\,,\label{eqmetric}
\end{equation}
\begin{equation}
 \square\varphi+\frac{16\pi\alpha}{\sqrt{-g}}\frac{\delta S_{c}}{\delta \varphi}=16\pi\int m'(\varphi) \frac{\delta^{(4)}(x-y(\tau))}{\sqrt{-g}}d\tau\,,\label{eq:pertphi}
\end{equation}
where
\begin{equation}
T_{{\rm scal}}^{\mu\nu}=\frac{1}{2}\partial^\mu\varphi\partial^\nu\varphi
-\frac{1}{4}g^{\mu\nu}g_{\alpha\beta}\partial^\alpha\varphi\partial^\beta\varphi,
\end{equation}
\begin{equation}
T_{c}^{\mu\nu}=-\frac{16\pi}{\sqrt{-g}}\frac{\delta S_{c}}{\delta g^{\mu\nu}},
\end{equation}
\begin{equation}
  T_p^{\mu\nu}=
  \int  m(\varphi)\frac{\delta^{(4)}(x-y(\tau))}{\sqrt{-g}}
  \frac{dy^\mu}{d\tau}\frac{dy^\nu}{d\tau} d\tau\,,
\end{equation}
$m'(\varphi)=dm(\varphi)/d\varphi$, $x$ describes the spacetime point and $y(\tau)$ is the particle's worldline.
We consider the inspiral of a small BH into a massive Schwarzschild BH with the extreme mass ratio between the small body and the central BH.
In theories like the scalar Gauss-Bonnet (sGB) gravity, the small BH carries a scalar charge.
If the conditions $\alpha T_{c}^{\mu\nu}\ll G^{\mu\nu}$ and $\alpha \frac{\delta S_{c}}{\delta \varphi}\ll \square\varphi$ are satisfied,
then Einstein's equation and  the scalar field equation for EMRIs reduce to \cite{Maselli:2020zgv}
\begin{align}
G^{\mu\nu}&\approx 8\pi T_p^{\mu\nu}=8\pi m_p\int \frac{\delta^{(4)}(x-y(\tau))}{\sqrt{-g}}
\frac{dy^\mu}{d\tau}\frac{dy^\nu}{d\tau} d\tau,\label{eq:pertG}\\
  \square\varphi&=S=4\pi m_p\,d\int \frac{\delta^{(4)}(x-y(\tau))}{\sqrt{-g}}d\tau\,,
\end{align}
where $m_p=m(\varphi_0)$, $d=4 m'(\varphi_0)/m(\varphi_0)$ is the  dimensionless scalar charge of the small BH and $\varphi_0$ stands for the background value of the scalar field.
The primary BH provides an approximate Schwarzschild background.
In the Teukolsky formalism, the Newman-Penrose scalar $\Psi_4$ and the scalar perturbation $\varphi$ are decomposed into spherical harmonic components as follows
\begin{equation}
\begin{split}
\Psi_4&=\int_{-\infty}^{+\infty}d\omega\sum_{lm}r^{-4}R_{\omega lm}(r)~{_{-2}}Y_{lm}(\theta,\phi)e^{-i\omega t}, \\
\varphi&=\int_{-\infty}^{+\infty}d\omega\sum_{lm}r^{-1}X_{\omega lm}(r)~{_{0}}Y_{lm}(\theta,\phi)e^{-i\omega t},
\end{split}
\end{equation}
where ${_s}Y_{lm}(\theta,\phi)$ is the spin-weighted spherical harmonics.
The radial functions $R_{\omega lm}(r)$ and $X_{\omega lm}(r)$ satisfy the inhomogeneous Teukolsky equation
\begin{equation}\label{teukolsky}
\begin{split}
\left[r^2f\frac{d^2}{dr^2}-2(r-M)\frac{d}{dr}+U_1(r)\right]R_{\omega l m}(r)&=-T_{\omega l m}(r),  \\
\left[f^2\frac{d^2}{dr^2}+\frac{2Mf}{r^2}\frac{d}{dr}+U_2(r)\right]X_{\omega l m}(r)&=-S_{\omega l m}(r),
\end{split}
\end{equation}
where $f=1-2M/r$, $M$ is the mass of the central BH and
\begin{equation}
\begin{split}
U_1(r)&=f^{-1}\left(\omega^2r^2-4i\omega(r-3M)\right)-(l-1)(l+2),\\
U_2(r)&=\omega^2-r^{-2}f\left(l(l+1)+\frac{2M}{r} \right).
\end{split}
\end{equation}
The source term in Eq.~\eqref{teukolsky} is
\begin{equation}
S_{\omega l m}(r)=\frac{1}{2\pi}\int dt d\Omega~ rfS~ {_{0}}\bar{Y}_{l,m}(\theta,\phi)e^{i\omega t},
\end{equation}
S will be given below and $T_{\omega l m}$ is given in \cite{Cutler:1994pb}.
The homogeneous Teukolsky equation admits two linearly independent solutions $X^{\text{in}}_{lm\omega}$ and $X^{\text{up}}_{lm\omega}$, with the following asymptotic values at the horizon $r=2M$ and at infinity,
\begin{equation}
X^{\text{in}}_{lm\omega}=
\begin{cases}
e^{-i\omega r^*},&\qquad (r^*\to-\infty)\\
A_{\text{out}}e^{i\omega r^*}+A_{\text{in}}e^{-i\omega r^*}, &\qquad (r^*\to+\infty)
\end{cases}
\end{equation}
\begin{equation}
X^{\text{up}}_{lm\omega}=
\begin{cases}
B_{\text{up}}e^{i\omega r^*}+B_{\text{in}}e^{-i\omega r^*},&\qquad (r^*\to-\infty)\\
e^{i\omega r^*},&\qquad (r^*\to+\infty)\\
\end{cases}
\end{equation}
where the tortoise radius $r^*=r+2M\ln{\left(r/2M-1\right)}$.
The solutions $X^{\text{in}}_{lm\omega}$ and $X^{\text{up}}_{lm\omega}$  are purely outgoing at infinity and purely ingoing at the horizon.
With the help of these homogeneous solutions, the solution to Eq.~\eqref{teukolsky} is
\begin{equation}
X(r)=\frac{X^{\text{in}}_{lm\omega}\int_{r^*}^{+\infty}S_{lm\omega}X^{\text{up}}_{lm\omega}dr^*+X^{\text{up}}_{lm\omega}\int_{-\infty}^{r^*}S_{lm\omega}X^{\text{in}}_{lm\omega}dr^*}{W},
\end{equation}
where the Wronskian of $X^{\text{in}}_{lm\omega}$ and $X^{\text{up}}_{lm\omega}$ is
\begin{equation}
W=X^{\text{in}}_{lm\omega}\frac{dX^{\text{up}}_{lm\omega}}{dr^*}-X^{\text{up}}_{lm\omega}\frac{dX^{\text{in}}_{lm\omega}}{dr^*}=2i\omega A_{\text{in}}.
\end{equation}

\section{Orbital evolution and waveform}
\label{sec3}
The timelike geodesics of the Schwarzschild spacetime on the equatorial plane $\theta=\pi/2$ are described by the equations
\begin{equation}
\label{orbite}
\begin{split}
   dt/d\tau &= \frac{E}{f(r)},\\
(dr/d\tau)^2 &= E^2-V ,\\
d\phi/d\tau &=\frac{L}{r^2},
\end{split}
\end{equation}
where $\{t,r,\theta,\phi\}$ are the usual Schwarzschild coordinates,
the constants $E$ and $L$ are the energy and angular momentum per unit mass, respectively.
The effective potential for radial motion is given by
\begin{equation}
V=f\left(1+\frac{L^2}{r^2}\right).
\end{equation}
To exploit the analogy with Keplerian orbits, it is useful to make the substitution
\begin{equation}
    r(\chi)=\frac{pM}{1+e\cos\chi}.
\end{equation}
The parameter $\chi$ ranges from 0 to $2\pi$ as $r$ goes from the periastron $r_1$ to the apastron $r_2$ and back to the periastron $r_1$;
The constants $p$ and $e$ can be derived from $E$ and $L$ by letting $dr/d\tau=0$ at the periastron and the apastron in Eq.~\eqref{orbite},
which leads to
\begin{equation}
\label{EL}
    \begin{split}
E^2&=E(p,e)^2=\frac{(p-2-2e)(p-2+2e)}{p(p-3-e^2)},\\
L^2&=L(p,e)^2=\frac{p^2M^2}{p-3-e^2}.
    \end{split}
\end{equation}
Integrating Eq.~\eqref{orbite} gives \cite{Cutler:1994pb}
\begin{equation}
  \begin{split}
t(\chi)&=p^2M(p-2-2e)^{1/2}(p-2+2e)^{1/2}\\
&\times \int_0^\chi d\chi'(p-2-2e\cos\chi')^{-1}(1+e\cos\chi')^{-2}\\
&\times (p-6-2e\cos\chi')^{-1/2},
  \end{split}
\end{equation}
and
\begin{equation}
\phi(\chi)=p^{1/2}\int_0^\chi\frac{d\chi'}{(p-6-2e\cos\chi')^{1/2}}.
\end{equation}
The radial period is given by $P=2t(\pi)$. The radial frequency and
the azimuthal frequency are given by $\Omega_r=2\pi/P$ and $\Omega_\phi=\phi(2\pi)/P$, respectively.
 We now proceed with the calculation of the source term in Eqs.~\eqref{teukolsky},
 taking the particle's worldline to be a bound geodesic in the Schwarzschild spacetime.
 After integration, the particle's stress-energy tensor becomes
 \begin{equation}
     \begin{split}
T^{\alpha\beta}_p(x)&=m_p \frac{u^{\alpha}u^{\beta}}{r'^2u^t}\delta(r-r')\delta(\cos\theta)\delta(\phi-\phi'),\\
S(x)&=\mu_s \frac{1}{r'^2u^t}\delta(r-r')\delta(\cos\theta)\delta(\phi-\phi').
     \end{split}
 \end{equation}
Here $\mu_s=4\pi m_{\rm p}\,d$ and $\{t,r',\pi/2,\phi'(t)\}$ describe the particle's worldline;
the four-velocity $u^\alpha=dx'^\alpha/d\tau$ can be obtained from Eq.~\eqref{orbite}.
Following the procedure given in \cite{Cutler:1994pb} we get
\begin{equation}
\begin{split}
S_{lm\omega}(r)&=\frac{\mu_s}{2\pi}~{_{0}}Y_{lm}(\frac{\pi}{2},0)\int_{-\infty}^{+\infty} dt~\frac{r'f}{r'^2u^t} \delta(r-r')e^{i(\omega t-m\phi')}\\
&=\mu_s~ {_{0}}Y_{lm}(\frac{\pi}{2},0) P^{-1}\sum_k\delta(\omega-\omega_{mk})\\
&\qquad\times\int_0^Pdt~\frac{r'f}{r'^2u^t} \delta(r-r')e^{i(\omega_{mk} t-m\phi')}\\
&=\mu_s~ {_{0}}Y_{lm}(\frac{\pi}{2},0)\Theta(r-r_1)\Theta(r_2-r)\\
&\qquad \times f^2 P^{-1}(E^2-V)^{-1/2}\sum_k\delta(\omega-\omega_{mk})\\
&\qquad \times\sum_{\pm}            \frac{1}{rf}               e^{\pm i[\omega_{mk}t(r)-m\phi(r)]},
\end{split}
\end{equation}
where $\omega_{mk}=m\Omega_\phi+k\Omega_r$.
Therefore, the solutions for scalar field at infinity and at the horizon are given by
\begin{equation}
\begin{split}
X_{lm\omega}^{\infty, H}(r)&\sim\frac{e^{\pm i\omega r^*}}{2i\omega A^{\text{in}}_{lm\omega}}\int_{-\infty}^{+\infty}dr^{*'}X^{\text{in},\text{up}}_{lm\omega}(r')S_{lm\omega}(r')\\
&\equiv \tilde{Z}_{lm\omega}^{\infty, H}e^{\pm i\omega r^*}.
\end{split}
\end{equation}
So we can get
\begin{equation}
\begin{split}
\tilde{Z}_{lm\omega}^{\infty,H}=&\frac{1}{2i\omega A^{\text{in}}_{lm\omega}}\int_{-\infty}^{+\infty}dr^{*'}X^{\text{in},\text{up}}_{lm\omega}(r')S_{lm\omega}(r')\\
=&\frac{\mu_s P^{-1}}{2i\omega A^{\text{in}}_{lm\omega}}~ {_{0}}Y_{lm}(\frac{\pi}{2},0)\sum_{\pm}\sum_k\delta(\omega-\omega_{mk})\int_{r_1}^{r_2}dr\frac{X^{\text{in},\text{up}}_{lm\omega}(r)}{r\sqrt{E^2-V}} e^{\pm i[\omega_{mk}t(r)-m\phi(r)]}\\
=&\mu_s\sum_kZ_{lmk}^{\infty,H}\delta(\omega-\omega_{mk}).
\end{split}
\end{equation}
The formula for $dE_{\text{scal}}^{\infty,H}/dt$ and $dL_{\text{scal}}^{\infty,H}/dt$ caused by the scalar field at infinity and at the horizon are given by
\begin{equation}
\label{energyadd}
\begin{split}
\frac{dE_{\text{scal}}^{\infty,H}}{dt}&=\frac{\mu_s^2}{4\pi}\sum_{lmk}\omega_{mk}^2|Z_{lmk}^{\infty,H}|^2=4\pi d^2 m_p^2 \sum_{lmk}\omega_{mk}^2|Z_{lmk}^{\infty,H}|^2 ,\\
\frac{dL_{\text{scal}}^{\infty,H}}{dt}&=\frac{\mu_s^2}{4\pi}\sum_{lmk}m\,\omega_{mk} |Z_{lmk}^{\infty,H}|^2=4\pi d^2 m_p^2 \sum_{lmk}m\,\omega_{mk} |Z_{lmk}^{\infty,H}|^2.
\end{split}
\end{equation}
The formula for $dE_{\text{grav}}^{\infty,H}/dt$ and $dL_{\text{grav}}^{\infty,H}/dt$ caused by the tensor field are given in \cite{Cutler:1994pb}.
The total energy and angular momentum flux are
\begin{equation}
\begin{split}
\label{tolflux}
\dot{E}_{\text{tot}}& =\dot{E}_{\text{grav}}+\dot{E}_{\text{scal}}=\dot{E}_{\text{grav}}^{\infty}+\dot{E}_{\text{grav}}^{H}+\dot{E}_{\text{scal}}^{\infty} +\dot{E}_{\text{scal}}^{H}  ,\\
\dot{L}_{\text{tot}}& =\dot{L}_{\text{grav}}+\dot{L}_{\text{scal}}=\dot{L}_{\text{grav}}^{\infty}+\dot{L}_{\text{grav}}^{H}+\dot{L}_{\text{scal}}^{\infty} +\dot{L}_{\text{scal}}^{H} .
\end{split}
\end{equation}

Our method to calculate the bound orbital evolution of the Schwarzschild spacetime caused by radiation reaction is as follows.
Based on the fact that the motion of the particle is geodesic over a time scale comparable to the orbital period,
we adopt the adiabatic approximation in which the radiation reaction operates over a much longer time scale.
The waves are generated by the orbiting particle, and the average is taken over several orbital periods.
The orbital parameters change as follows
\begin{equation}
\label{balance}
    \left\langle \frac{d E_{\text{tot}}}{dt} \right \rangle=\dot{E}_{\text{tot}}=-m_p\dot{E},\qquad \left\langle \frac{d L_{\text{tot}}}{dt} \right\rangle= \dot{L}_{\text{tot}}=-m_p\dot{L}.
\end{equation}
Using Eq.~\eqref{balance}, we can then infer the time-averaged rates of change of the orbital parameters.
Since $E$ and $L$ are functions of $p$ and $e$, we have
\begin{equation}
\label{balance2}
\begin{split}
   -\dot{E}_{\text{tot}}&=m_p\frac{\partial E}{\partial p}\frac{dp}{dt}+m_p\frac{\partial E}{\partial e}\frac{de}{dt},\\
-\dot{L}_{\text{tot}}&=m_p\frac{\partial L}{\partial p}\frac{dp}{dt}+m_p\frac{\partial L}{\partial e}\frac{de}{dt}.
\end{split}
\end{equation}
These equations can easily be inverted.
Using Eqs.~\eqref{EL} and \eqref{balance2}, we find
\begin{equation}
\label{p}
\begin{split}
m_p\frac{dp}{dt}=&\frac{2(p-3-e^2)^{1/2}}{(p-6-2e)(p-6+2e)}\left[
p^{3/2}(p-2-2e)^{1/2}(p-2+2e)^{1/2}\dot{E}_{\text{tot}}\right.\\
&\qquad\left.-(p-4)^2\dot{L}_{\text{tot}}/M\right]
\end{split}
\end{equation}
and
\begin{equation}
\label{e}
\begin{split}
m_p\frac{de}{dt}=&\frac{(p-3-e^2)^{1/2}}{ep(p-6-2e)(p-6+2e)}\left[
(1-e^2)\left(4e^2+(p-2)(p-6)  \right)\dot{L}_{\text{tot}}/M\right.\\
&\qquad \left.-p^{3/2}(p-6-2e^2)(p-2-2e)^{1/2}(p-2+2e)^{1/2}\dot{E}_{\text{tot}}\right] .
\end{split}
\end{equation}
It is important to notice that Eqs.~\eqref{p} and \eqref{e} are singular at $p=6+2e$, where the trajectory stops.

The orbit-averaged trajectory is determined over time in terms of $\{p(t), e(t), \Phi_{\varphi}(t)$, $\Phi_{r}(t)\}$.
The phases $\Phi_{\varphi,r}$ given above are the integral over time of the orbit's fundamental frequencies:
\begin{equation}
\Phi_{\varphi,r}=\int_0^tdt'\,\Omega_{\varphi,r}\left(p(t'),e(t')\right).
\end{equation}
EMRI waveforms are represented by the complex time-domain dimensionless strain
\begin{equation}
h_{+}-ih_{\times}=-\frac{2m_p}{d_L}\sum_{lmk}\frac{Z^{\infty}_{\text{grav},lmk}}{\omega_{mk}^2}~{_{-2}}Y_{lm}(\theta,0)e^{im\phi}e^{-i\Phi_{mk}(t)},
\end{equation}
where $Z^{\infty}_{\text{grav},lmk}$ is  the Teukolsky mode amplitude for tensor field $\Psi_4$ far away from the source \cite{Katz:2021yft},  $\theta$ is the source-frame polar viewing angle, $\phi$ is the source-frame azimuthal viewing angle, $d_L$ is the luminosity distance, and $\{l,m,k\}$ are the indices describing the frequency-domain harmonic mode decomposition.
The indices $l,m,k$ label the orbital angular momentum, azimuthal and radial modes, respectively.
$\Phi_{mk}(t)=m\Phi_{\varphi}+k\Phi_r$ is the summation of decomposed phases for each given mode.
To estimate the detectability of the scalar charge,
we introduce the noise-weighted inner product to define the faithfulness  between two templates
\begin{equation}
\langle s_1|s_2 \rangle=2\int_{f_{\rm min}}^{f_{\rm max}}\frac{\tilde{s_1}(f)\tilde{s_2}^*(f)+\tilde{s_1}^*(f)\tilde{s_2}(f)}{S_{n}(f)}df,
\end{equation}
$\tilde{s}_{1}(f)$ is the Fourier transform of the time domain signal,
and its complex conjugate is $\tilde{s}_{1}^{*}(f)$.
The faithfulness between two signals is defined as
\begin{equation}\label{eq:def_F}
\mathcal{F}[s_1,s_2]=\max_{\{t_c,\phi_c\}}\frac{\langle s_1\vert
	s_2\rangle}{\sqrt{\langle s_1\vert s_1\rangle\langle s_2\vert s_2\rangle}}\ ,
\end{equation}
where $(t_c,\phi_c)$ are time and phase offsets \cite{Lindblom:2008cm},
The signal and noise spectral density $S_n(f)$ for space-based detectors are given in \cite{Chua:2015mua, Katz:2021yft, Barsanti:2022ana}.

\section{Numerical results}
\label{sec4}
\begin{figure*}[htp]
  \centering
  \includegraphics[width=0.45\textwidth]{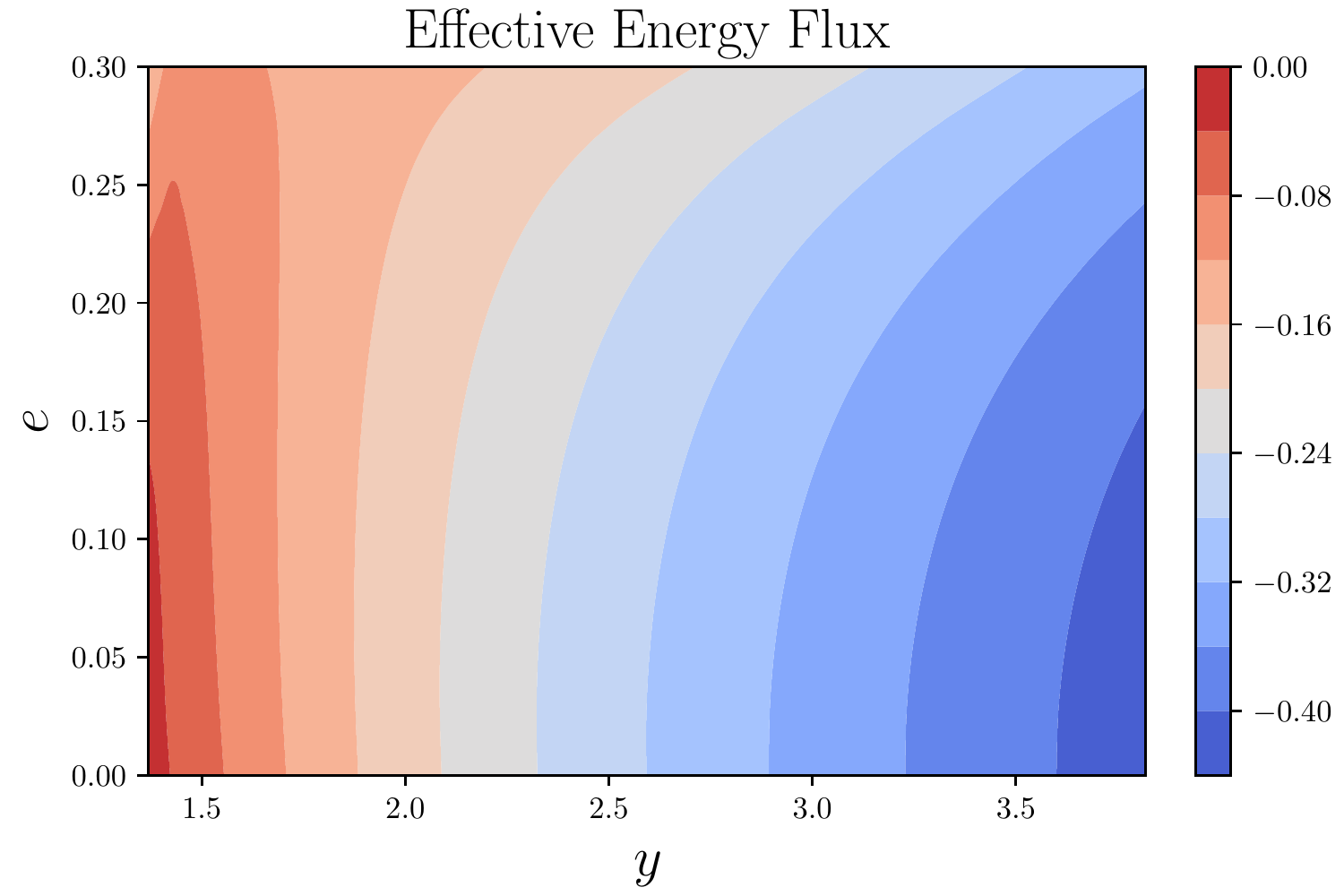}
  \includegraphics[width=0.45\textwidth]{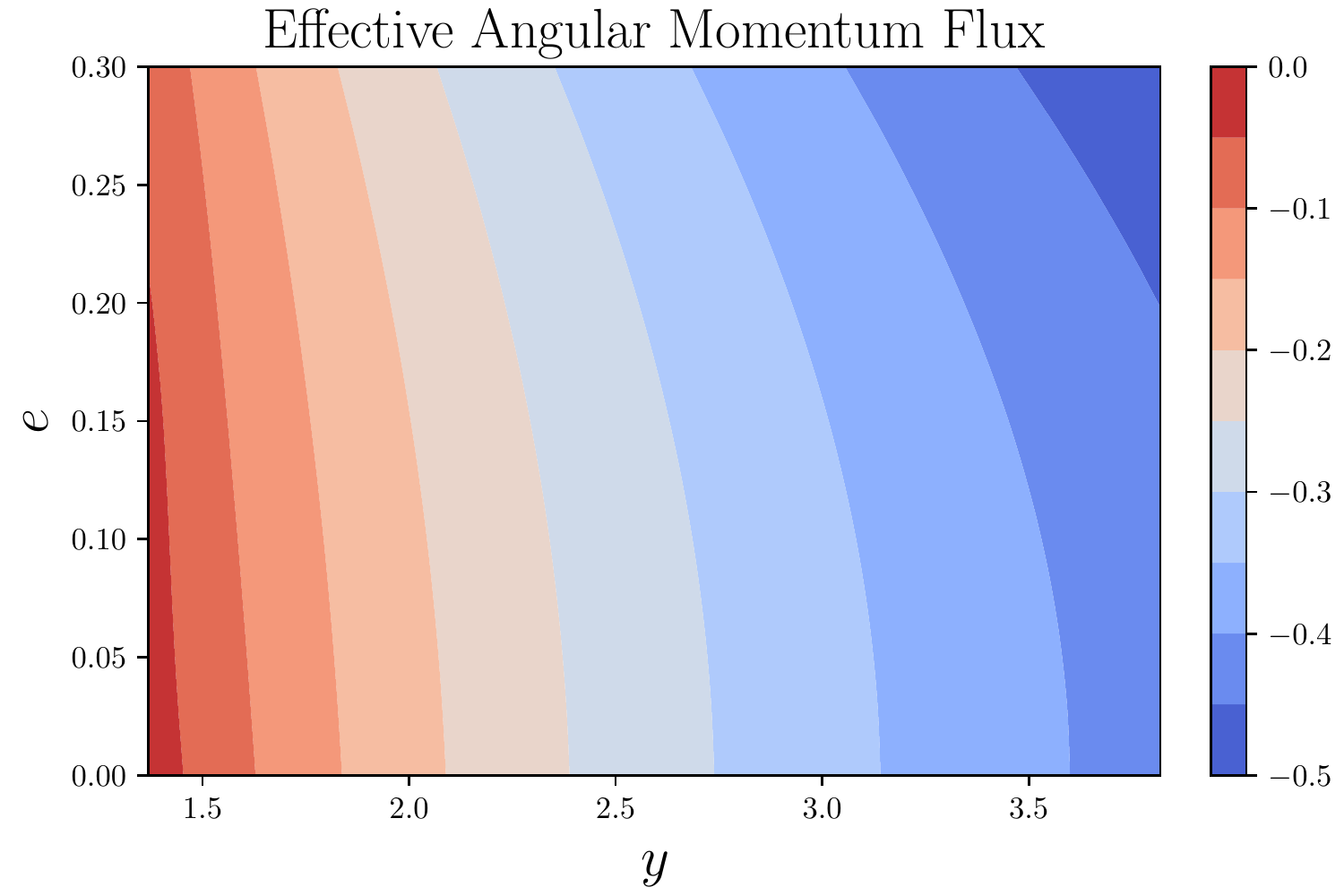}
  \caption{The effective energy flux and angular momentum flux for the scalar filed with $d=1$.
  The left panel shows the effective energy flux $\dot{E}_{\text{spl}}$ and the right panel shows the effective angular momentum flux $\dot{L}_{\text{spl}}$.
 }
  \label{eaflux}
\end{figure*}

As seen in Eq.~\eqref{energyadd}, the numerical calculation of $\dot{E}_{\text{tot}}$ and $\dot{L}_{\text{tot}}$  involves the truncation of infinite sums over $l$ and $k$.
This truncation obviously limits the accuracy of the numerical results.
We choose $l_{\text{max}}=6$, $k_{\text{min}}=-4$ and $k_{\text{max}}=10$ to reach the accuracy about $10^{-4}$ for eccentric orbits with low eccentricities $e\leq0.3$.
In Table~\ref{comparision} we compare our results on the radiated energy and angular momentum flux with those in Refs. \cite{Cutler:1994pb,Warburton:2011hp} for two representative points in the $p-e$ plane.
In order to efficiently interpolate the fluxes with bicubic splines, we introduce $y=\ln(p-p_s+3.9)$ \cite{Chua:2020stf} and $e$ to build a uniform grid in $(y,e)$ space with $1.37\leq y\leq3.82$ and $0.0\leq e \leq0.3$ in steps of $0.025$ where $p_s=6+2e$ \cite{Stein:2019buj}.
The large errors in the interpolation over the uniform grid of the actual flux values are intolerable.
For the point $(p,e)=(6.688, 0.3)$, the actual energy and angular fluxes from the numerical calculation are $2.359\times 10^{-4}$ and $3.301\times 10^{-3}$, the interpolated energy and angular momentum values are $2.450\times 10^{-4}$ and $3.377\times 10^{-3}$, the relative errors can reach $3.83\%$ and $2.31\%$.
Generally we would like to control the relative error under $\sim 1\%$ \cite{Cutler:1994pb}.
In order to reduce the error in the interpolation, we subtract out the leading PN behavior from the actual flux values and interpolate over the resulting effective fluxes
$\dot{E}_{\text{spl}}=(\dot{E}_{\text{scal}}-\dot{E}_{\text{PN}})\Omega_{\varphi}^{-10/3}$ and  $\dot{L}_{\text{spl}}=(\dot{L}_{\text{scal}}-\dot{L}_{\text{PN}})\Omega_{\varphi}^{-7/3}$.
The leading PN behaviors for scalar fields are given by \cite{Cardoso:2020iji}
\begin{equation}
\begin{split}
\dot{E}_{\text{PN}}&=\frac{1}{3}\frac{1+e^2/2}{(1-e^2)^{5/2}}\Omega_{\varphi}^{8/3},\\
\dot{L}_{\text{PN}}&=\frac{1}{3}\frac{1}{1-e^2}\Omega_{\varphi}^{5/3}.
\end{split}
\end{equation}
We construct bicubic splines over $(y,e)$ of $(\dot{E}_{\text{spl}},\dot{L}_{\text{spl}})$ and reconstruct the energy and angular momentum fluxes.
For the same point $(p,e)=(6.688, 0.3)$, the interpolated energy and angular momentum flux values are $2.346\times 10^{-4}$ and $3.290\times 10^{-3}$, the relative errors can reduce to $0.57\%$ and $0.31\%$, which satisfies our accuracy goal. 
The effective fluxes $(\dot{E}_{\text{spl}},\dot{L}_{\text{spl}})$ can be seen in Fig. \ref{eaflux}.
Making use of the FastEMRIWaveforms, based on the open code \cite{Katz:2021yft}, we can add the influence of the scalar field into the code by adding scalar flux values and get fast trajectory evolution and waveforms.

\begin{table}
  \centering
  	\begin{tabular}{|c|c|c|c|c|}
		\hline
$(p,e)$ & \multicolumn{2}{|c|}{(7.50478,\ 0.188917)} & \multicolumn{2}{|c|}{(10,\ 0.2)} \\ \hline
& This paper& Ref \cite{Cutler:1994pb}  & This paper& Ref \cite{Warburton:2011hp}\\ \hline
$M^2\dot{E}_{\text{grav}}^{\infty}/m^2_p$ & $3.16698\times10^{-4}$& $3.16804\times10^{-4}$ &/& / \\ \hline
$M^2\dot{E}_{\text{grav}}^{H}/m^2_p$ & $4.54079\times10^{-7}$& / & /& /                \\ \hline
$M^2\dot{E}_{\text{scal}}/m_p^2$ & /& / & $3.21274\times10^{-5}$& $3.21331\times10^{-5}$    \\ \hline
$M\dot{L}_{\text{grav}}^{\infty}/m^2_p$ & $5.96409\times10^{-3}$& $5.96562\times10^{-3}$ & /
&  /
\\ \hline
$M\dot{L}_{\text{grav}}^{H}/m^2_p$ & $7.62174\times10^{-6}$& / & /
& /
\\ \hline
$M\dot{L}_{\text{scal}}/m_p^2$ & /& / & $9.62487\times10^{-4}$& $9.62608\times10^{-4}$ \\ \hline
	\end{tabular}
    \caption{Comparison of the results on the energy and angular momentum fluxes in this paper and those in Refs. \cite{Cutler:1994pb,Warburton:2011hp} for two representative points in the $p-e$ plane.
    For a given $(p,e)$, we calculate the energy $\dot{E}_{\text{scal}}$ and angular momentum $\dot{L}_{\text{scal}}$ carried away by the scalar field with the scalar charge $d=1$, the energy $\dot{E}_{\text{grav}}^{\infty,H}$ and angular momentum $\dot{L}_{\text{grav}}^{\infty,H}$ carried away by the tensor field at infinity and at the horizon, respectively.
    }
    \label{comparision}
\end{table}

\begin{figure*}[htp]
  \centering
  \includegraphics[width=0.85\textwidth]{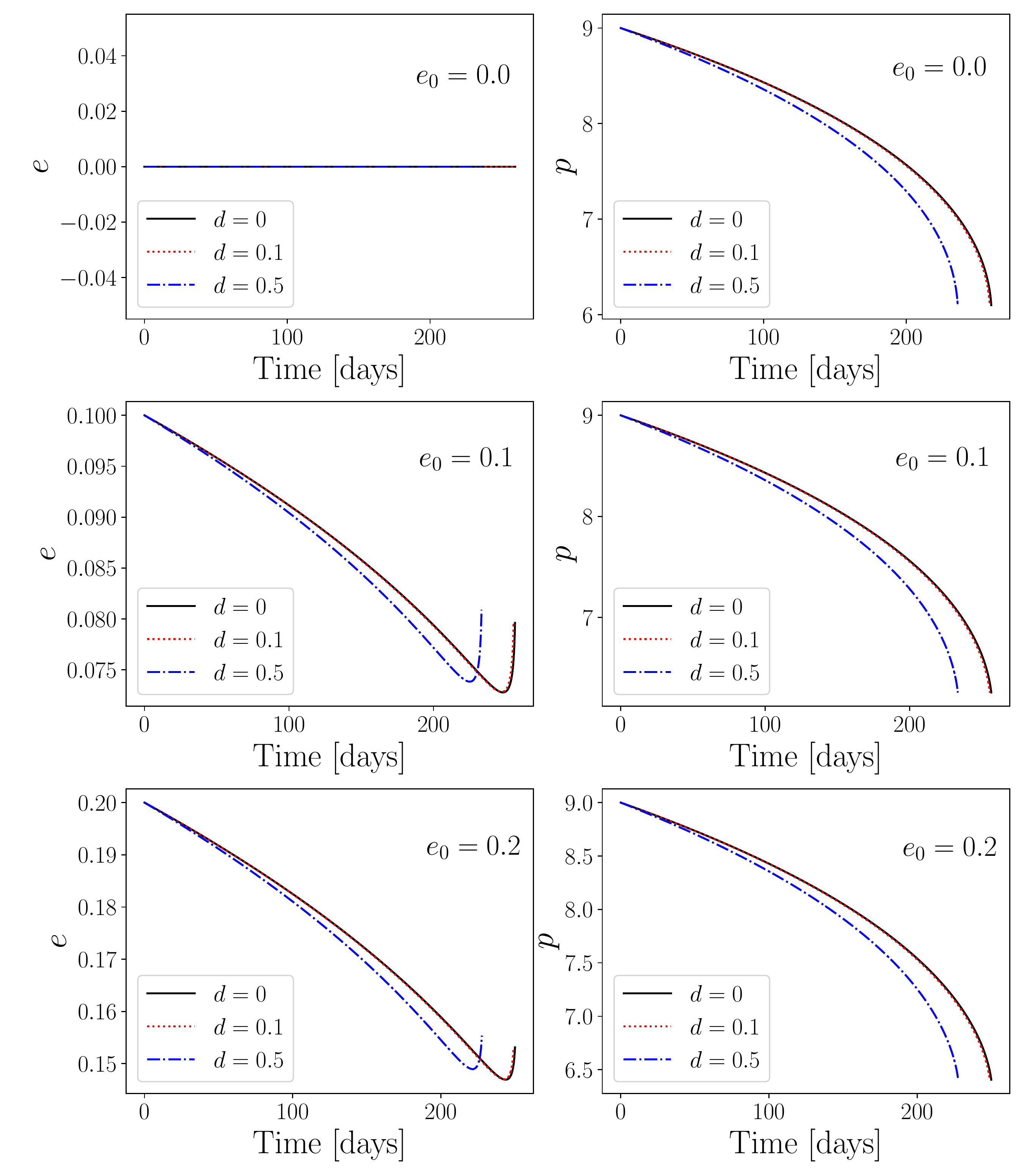}
  \caption{The orbital evolution for different values of the scalar charge. The initial conditions are chosen as $(p_0, e_0)$= $(9.0,0.0)$, $(9.0, 0.1)$ and $(9.0, 0.2)$, respectively.
  The left panels show the evolution of the eccentricity $e$ with time, and the right panels show the evolution of $p$ with time.}
  \label{orbit}
\end{figure*}

\begin{figure*}[htp]
  \centering
  \includegraphics[width=0.9\textwidth]{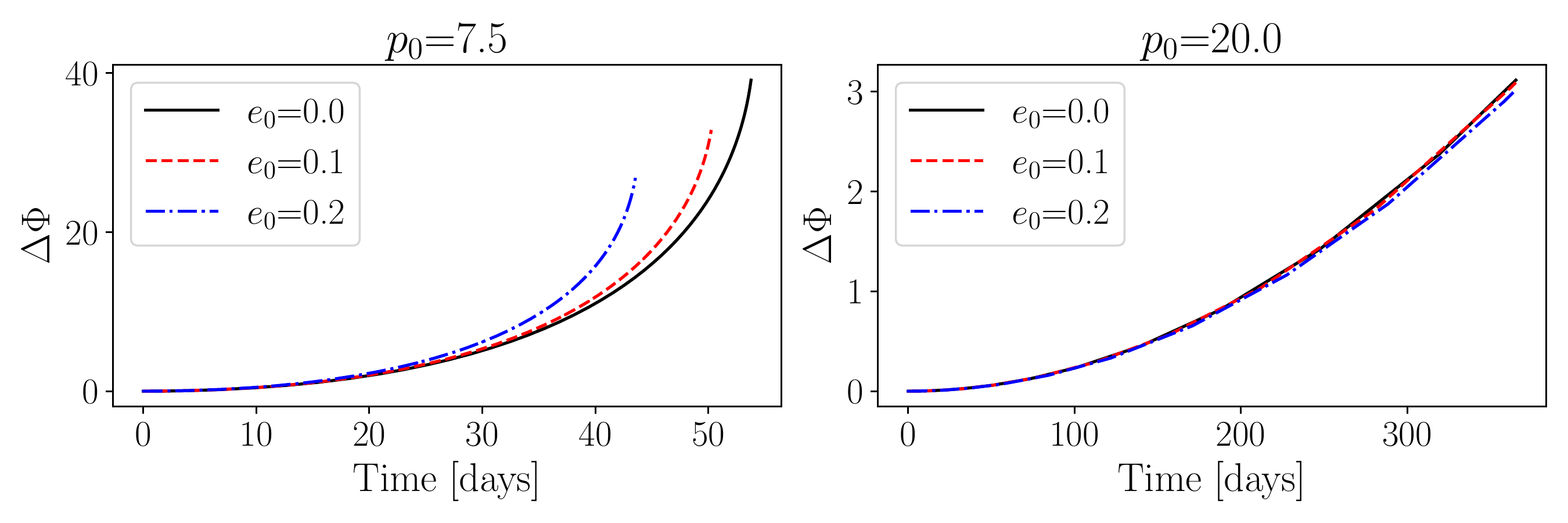}
  \caption{The accumulated orbital phase difference $\Delta\Phi=\Phi_\varphi(d=0.1)-\Phi_\varphi(d=0)$ as a function of time for different eccentric orbits. The initial values of the semi-latus rectum are $p_0=7.5$ and $p_0=20$ in the left and right panels, respectively.}
  \label{deltaphi1}
\end{figure*}

\begin{figure*}[htp]
  \centering
  \includegraphics[width=0.9\textwidth]{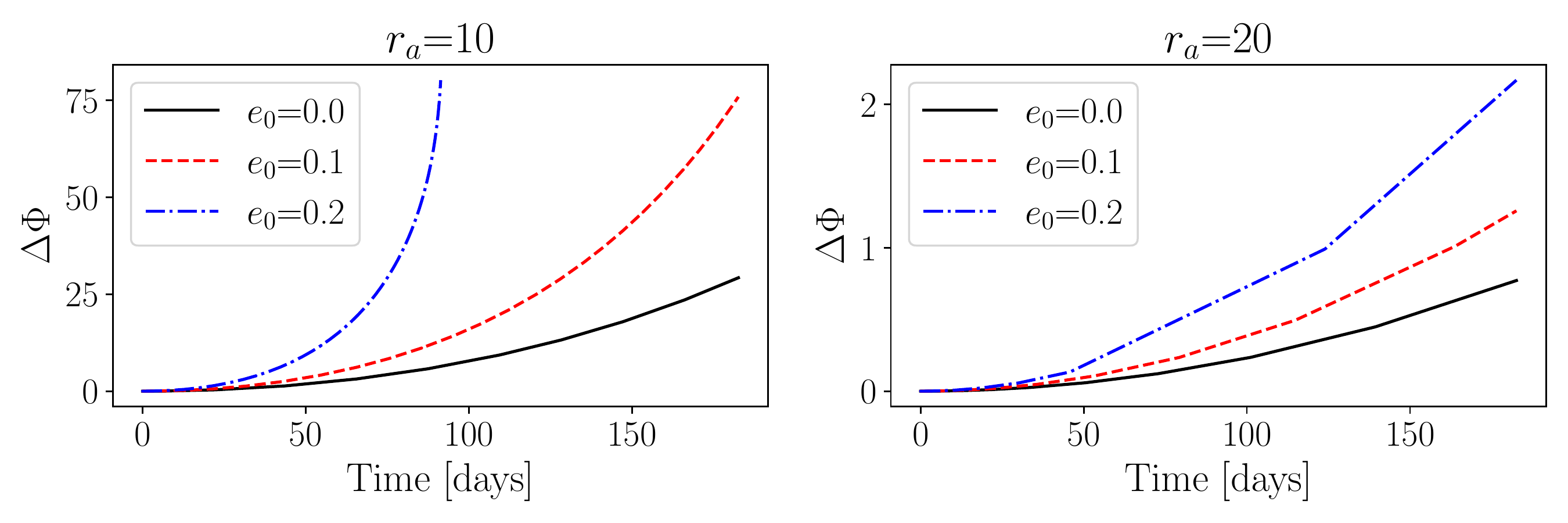}
  \caption{The accumulated orbital phase difference $\Delta\Phi=\Phi_\varphi(d=0.1)-\Phi_\varphi(d=0)$ as a function of time for different eccentric orbits. The initial values of the apastron are $r_a=10$ and $r_a=20$ in the left and right panels, respectively.}
  \label{deltaphi2}
\end{figure*}

\begin{figure*}[htp]
  \centering
  \includegraphics[width=0.45\textwidth]{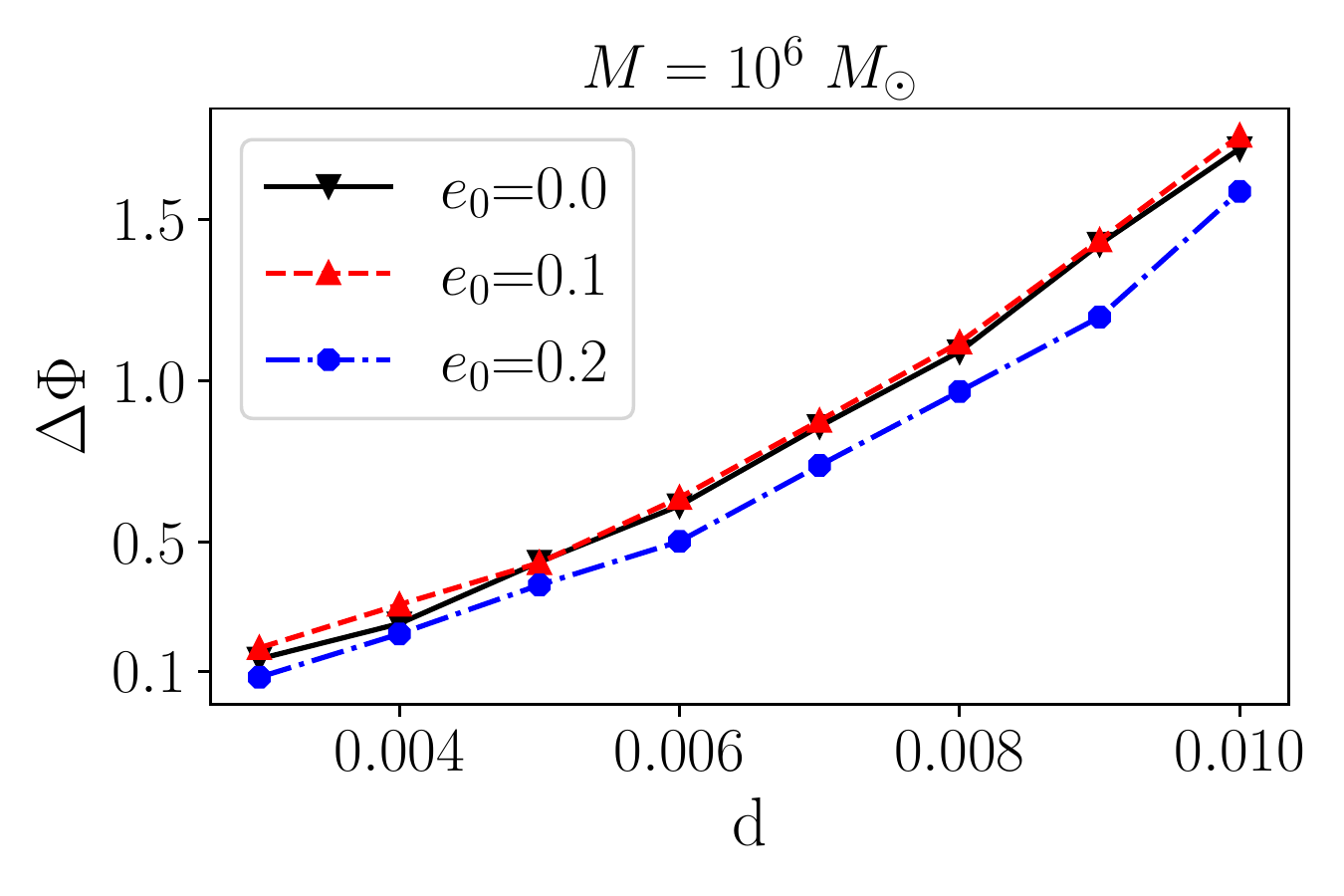} \quad
    \includegraphics[width=0.45\textwidth]{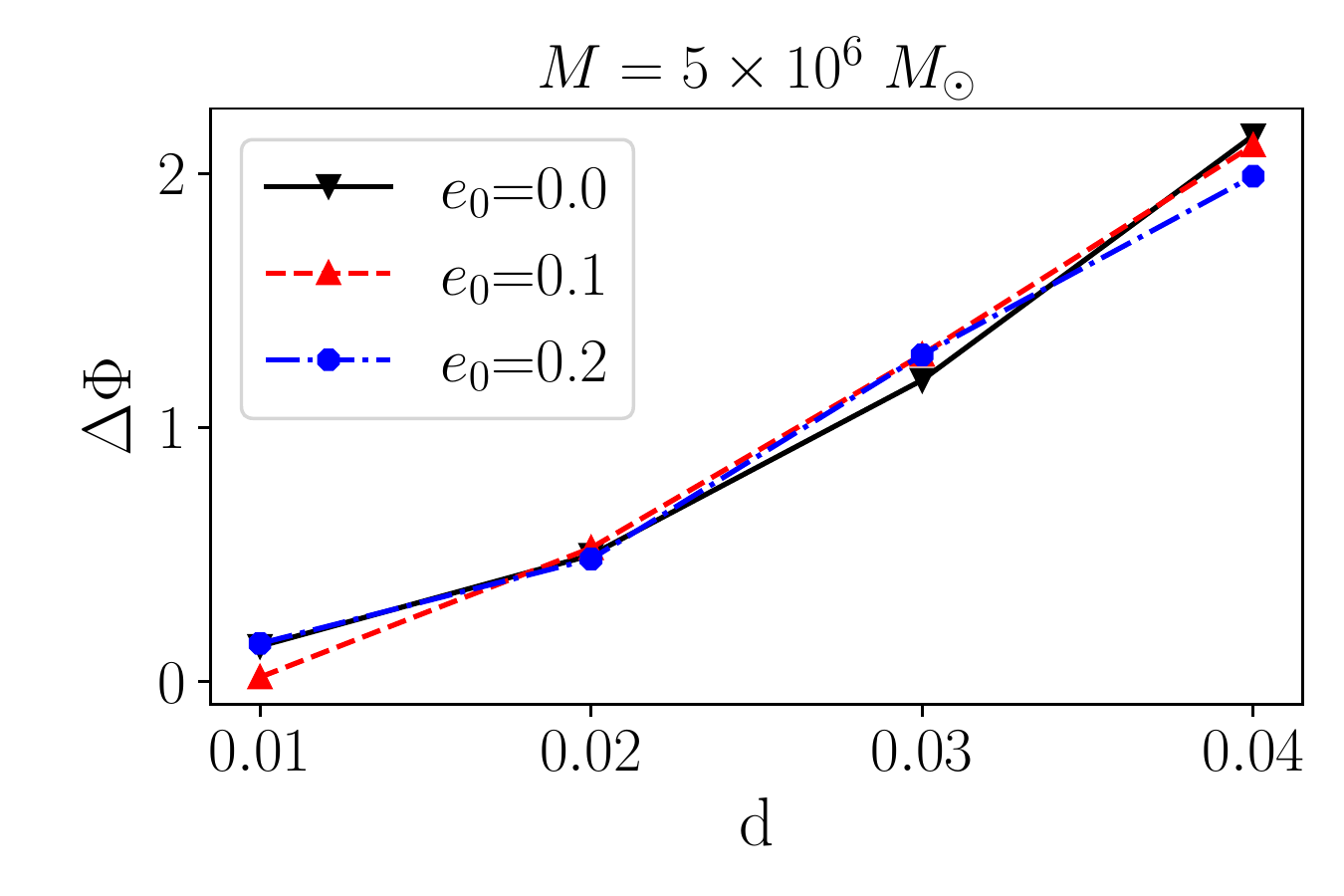}
  \caption{The accumulated orbital phase difference $\Delta\Phi=|\Phi_\varphi(d)-\Phi_\varphi(d=0)|$ versus the scalar charge for different initial values of the eccentricity $e_0$ and different masses for the central BH of EMRIs.
  $e_0$ is the initial eccentricity that inspiral starts and $p_0$ is adjusted to ensure one-year observation before the merger.
  The left panel represents the EMRIs with $M=10^6~M_{\odot}$ and $m_p=10~M_{\odot}$ and the right panel represents the EMRIs with $M=5\times 10^6~M_{\odot}$ and $m_p=10~M_{\odot}$.
}
  \label{dephasing}
\end{figure*}

We choose EMRI systems with $M=10^6~M_\odot$ for the central BH and $m_p=10~M_\odot$ for the small BH.
The orbital evolution from different initial conditions with various scalar charges is shown in Fig.~\ref{orbit}.
We find that the orbital parameter $p$ always decreases with time and the rate of change is faster for larger scalar charge $d$.
Scalar fields accelerate the evolution of $p$ since the scalar fields carry away additional energy and angular momentum to make the particle fall into the central BH faster.
For the orbital parameter $e$, the eccentricity decreases first and then increases near $p_s=6+2e$.
This change of eccentricity $e$ can be interpreted as a precursor effect to the eventual plunging of the orbit caused by the strong field effect, as discussed in \cite{Cutler:1994pb}.
Furthermore, we find that the time it takes $e$ to reach the minimum is shorter and the values of $p$ and $e$ at the turning point when $e$ reaches the minimum are bigger for larger scalar charge $d$.
The reason is that both the scalar and tensor fields enhance the gravitational field
so that the strong field effect comes into play further away from the central BH.
It's obvious from Eq.~\eqref{energyadd} that the scalar field with a bigger scalar charge $d$ carries away more energy and angular momentum to make the evolution of parameters faster.
Given a scalar charge, we can quantify the GR deviation caused by the scalar field with the scalar charge through the number of cycles accumulated after long-time evolution.
Fig. \ref{deltaphi1} shows the accumulated dephasing $\Delta\Phi=\Phi_\varphi(d)-\Phi_\varphi(d=0)$
with time for different initial values of $p_0$ and $e_0$.
Fig. \ref{deltaphi2} shows the accumulated dephasing $\Delta\Phi=\Phi_\varphi(d)-\Phi_\varphi(d=0)$
with time for different initial values of $r_a$ and $e_0$ as discussed in \cite{Barsanti:2022ana},
here $r_a=p_0/(1-e_0)$ is the apastron.
Starting from a smaller orbital distance with $p_0\leq 8$,
the accumulated dephasing $\Delta\Phi$ is bigger for larger initial eccentricity $e_0$
for the same evolution time.
However, starting from a larger orbital distance with $p\ge 20$,
the accumulated dephasing $\Delta\Phi$ is smaller for larger initial eccentricity $e_0$
for the same evolution time.
If we use different initial values of the apastron $r_a$,
then we find that the accumulated dephasing $\Delta\Phi$
is always bigger for larger initial eccentricity $e_0$ regardless of the initial values of $r_a$.
This result shows different behaviors between weak and strong fields for eccentric orbits and we will explain it in the next section.
In Fig. \ref{dephasing} we show the dephasing $\Delta\Phi$ as a function of the scalar charge $d$ for EMRIs with different eccentricities and different masses for the central BH.
The results show that bigger values of the scalar charge lead to larger dephasing,
and the dephasing decreases with a larger mass for the central BH and a larger initial eccentricity.


\begin{figure*}[htp]
  \centering
\includegraphics[width=0.95\textwidth]{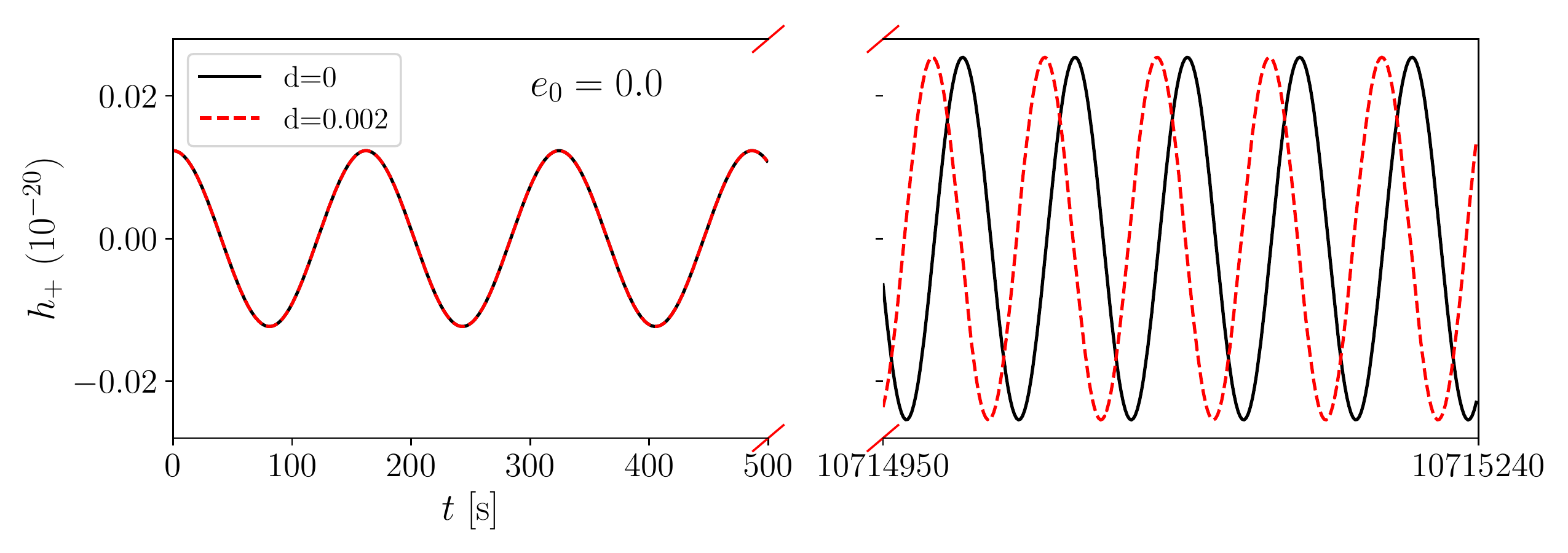}\\
\includegraphics[width=0.95\textwidth]{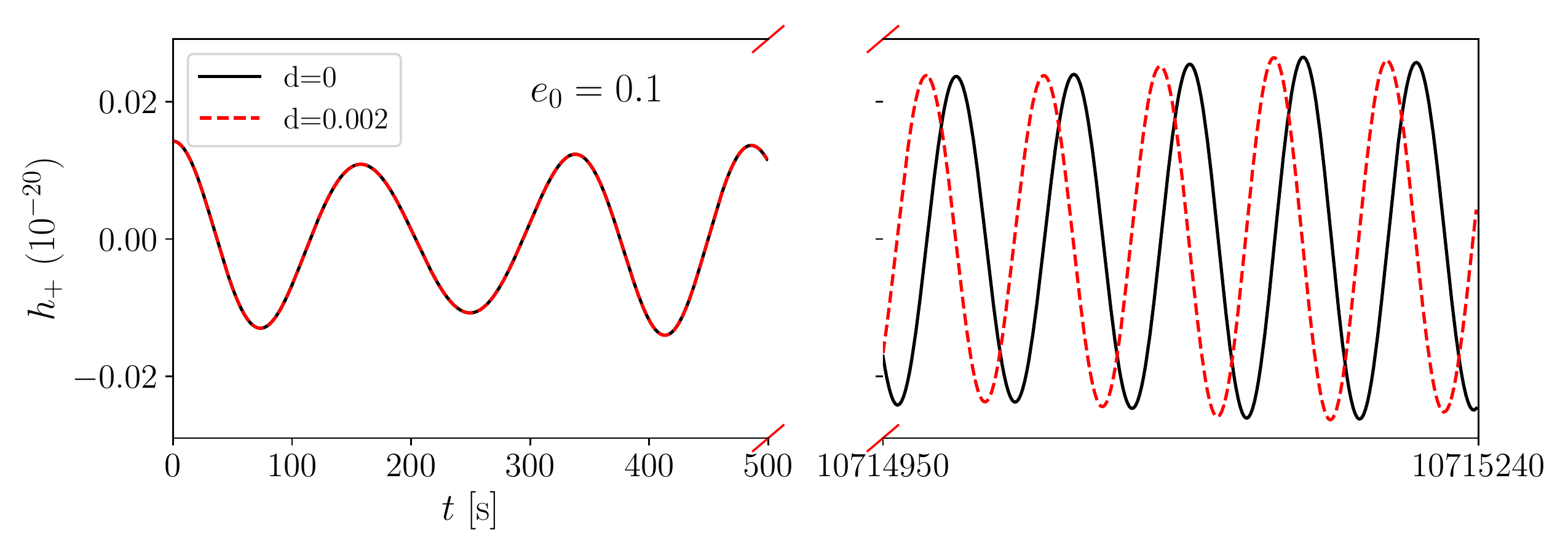}\\
\includegraphics[width=0.95\textwidth]{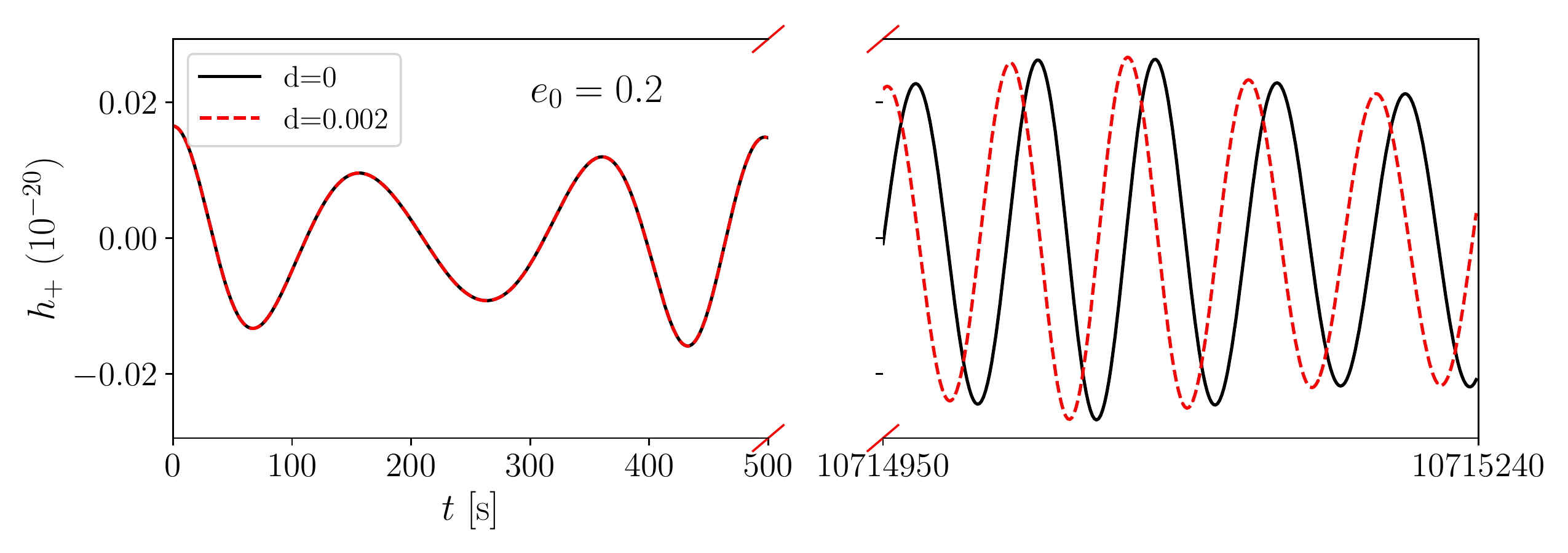}
  \caption{The waveforms with different scalar charge $d$ for different initial values of the eccentricity $e_0$ after a long-time evolution.
Here we take the initial semi-latus rectum $p_0=14$, and the luminosity distance $d_L=1$ Gpc.}
  \label{waveform}
\end{figure*}

\begin{figure*}[htp]
  \centering
  \includegraphics[width=0.8\textwidth]{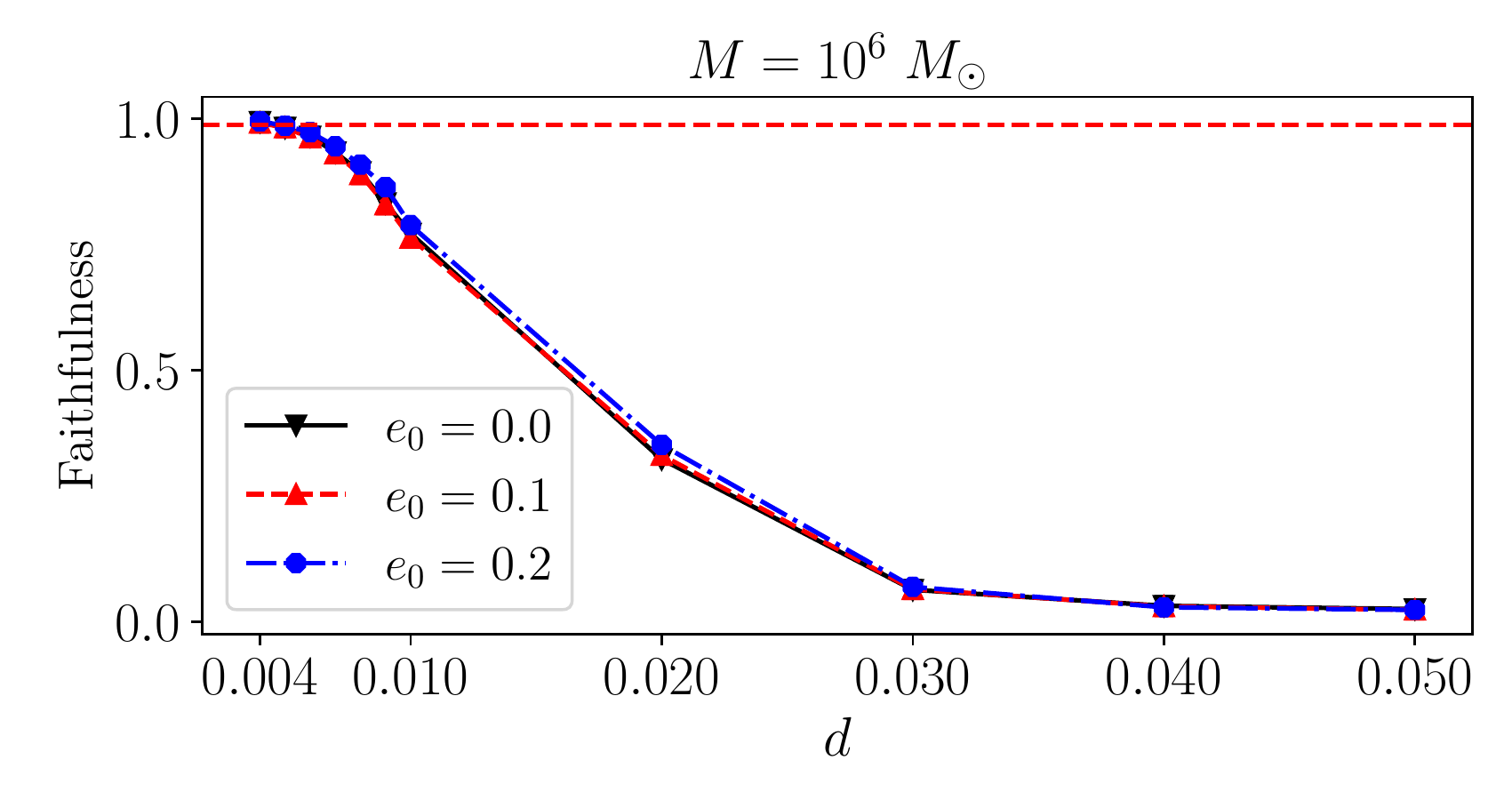}
  \caption{The faithfulness of GW signals between the case with and without the scalar charge as a function of the scalar charge $d$ when $M=10^6~M_{\odot}$. $e_0$ is the initial eccentricity that inspiral starts, and $p_0$ is adjusted to ensure one-year observation before the merger.
  The horizontal dashed line represents the detection limit $\mathcal{F}_n\leq 0.988$ with LISA.}
  \label{faithfulness1}
\end{figure*}

\begin{figure*}[htp]
  \centering
  \includegraphics[width=0.8\textwidth]{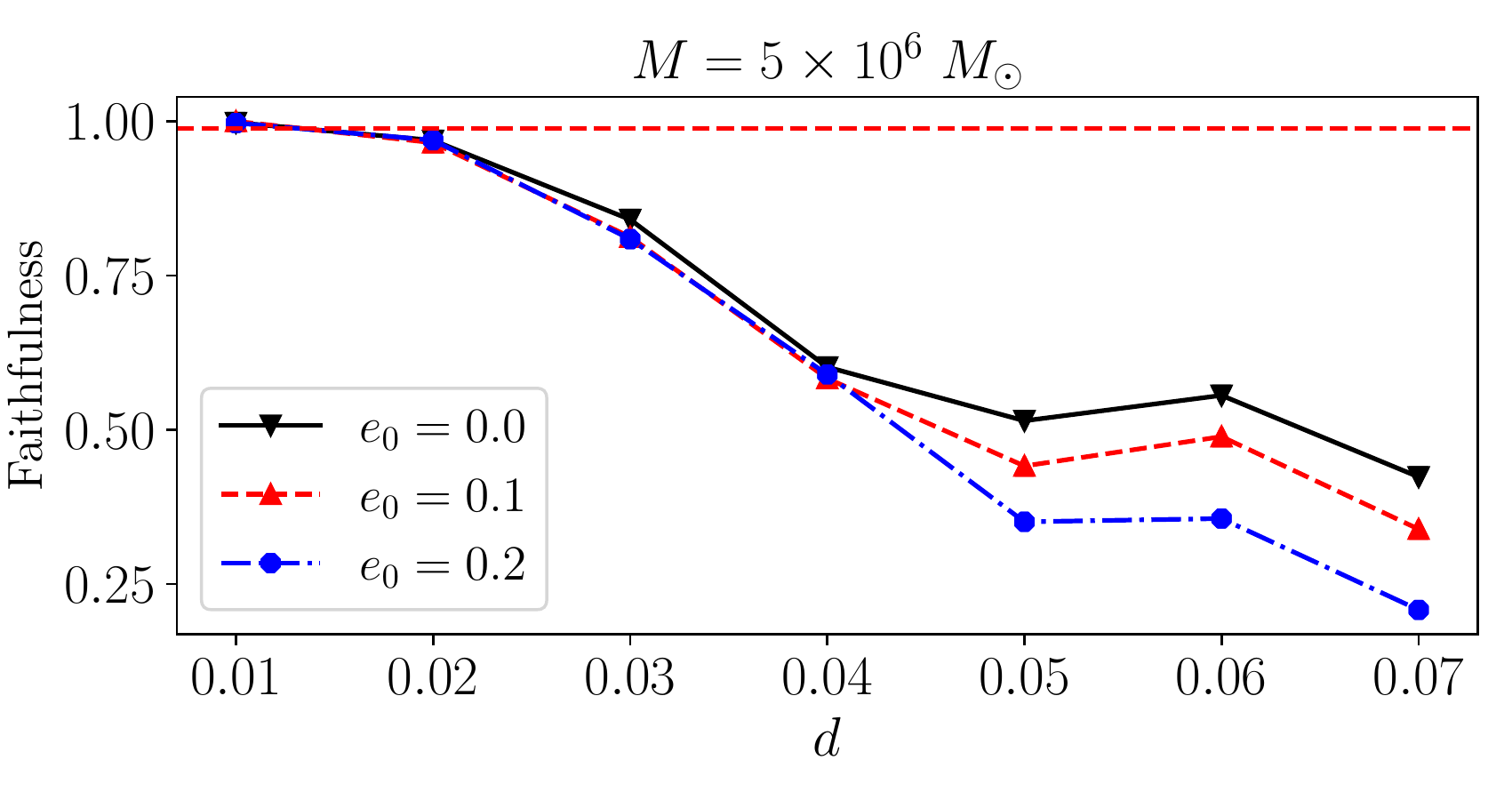}
  \caption{The faithfulness of GW signals between the case with and without the scalar charge as a function of the scalar charge $d$ when $M=5\times10^6~M_{\odot}$. $e_0$ is the initial eccentricity that inspiral starts, and $p_0$ is adjusted to ensure one-year observation before the merger.
  The horizontal dashed line represents the detection limit $\mathcal{F}_n\leq 0.988$ with LISA.}
  \label{faithfulness2}
\end{figure*}

It's obvious that larger $\Delta\Phi$ helps LISA detect the scalar charge easier for quasicircular orbits.
However, for eccentric orbits the ability of distinguishing modified gravity from GR not only depends on $\Delta\Phi$ but also depends on the eccentricity $e$.
For the convenience of discussion, 
we assume that starting at $t=0$ the initial phase and  positions of the small BH are the same for the situation with and without scalar charges.
Fig.~\ref{waveform} shows the GW waveforms with the scalar charge $d=0.002$ and $d=0$ after a long-time evolution from the initial position.
We see that as $\Delta\Phi$ accumulates, the dephasing between the waveforms becomes apparent at the time $t=10714950$ s.
The amplitude shapes of GWs are different for different eccentricities and the waveforms become sharper for higher eccentricities.
It was argued in Ref. \cite{Chatziioannou:2017tdw} that
two signals can be distinguished by LISA if $\mathcal{F}_n\leq0.988$.
In Figs. \ref{faithfulness1} and \ref{faithfulness2}, we show
the faithfulness between two GW signals with and without the scalar charge as a function of the scalar charge $d$ for different eccentric orbits.
These results suggest that LISA can detect the scalar charge around $d\ge 0.005$
through one-year observation of EMRIs consisting of a central BH with the mass $M=10^6~M_{\odot}$ and a small BH with the mass $m_p=10~M_{\odot}$.
For EMRIs with $M=5\times10^6~M_{\odot}$ and $m_p=10~M_{\odot}$,
the detection limit for the scalar charge with LISA is $d\ge 0.02$.
From Figs. \ref{faithfulness1} and \ref{faithfulness2},
we see that the effect of the scalar charge is more significant for eccentric inspirals than circular inspirals,
$d$ is correlated with $M$ and is anti-correlated with $e_0$.


\section{Analytical results}
\label{sec5}
In order to understand the strange behaviors of $\Delta\Phi$ for eccentric orbits with different initial values of $p_0$ and $e_0$ as shown in Fig. \ref{deltaphi1} and the numerical results presented in Ref. \cite{Barsanti:2022ana}, we calculate both scalar and gravitational energy fluxes emitted to infinity for eccentric orbits.
We expand the ingoing wave function $X_{\omega lm}^{\text{in}}$  in terms of small parameters $z=\omega r\sim v$ and $\epsilon=2M\omega \sim v^3$ \cite{Ohashi:1996uz},
\begin{equation*}
X_{\omega 0m}^{\text{in}}=z-\frac{z^3}{6}+\epsilon\left(\frac{-5z^2}{6}+\frac{17z^4}{180}\right),
\end{equation*}
\begin{equation*}
X_{\omega 1m}^{\text{in}}=\frac{z^2}{3}-\frac{z^4}{30}+\frac{z^6}{840}+\epsilon\left(\frac{-z}{6}-\frac{-7z^3}{60}+\frac{151z^5}{15120}\right), 
\end{equation*}
\begin{equation*}
X_{\omega 2m}^{\text{in}}=\frac{z^3}{15}-\frac{z^5}{210}+\epsilon\left(\frac{-z^2}{15}-\frac{4z^4}{315}+\frac{z^6}{1080}\right),
\end{equation*}
\begin{equation}
X_{\omega 3m}^{\text{in}}=\frac{z^4}{105}.    
\end{equation}
For $A^{\text{in}}_{lm \omega}$, we obtain \cite{Ohashi:1996uz}
\begin{equation}
A^{\text{in}}_{lm \omega}=\frac{1}{2}i^{l+1}e^{-i\epsilon \ln \epsilon}\left[1+\epsilon\left(p^{(1)}_{lm}+iq^{(1)}_{lm}\right)+\cdots \right],
\end{equation}
where
\begin{equation}
\begin{split}
p^{(1)}_{lm}&=-\frac{\pi}{2},\\
q^{(1)}_{lm}&=\frac{1}{2}\left[\psi(l)+\psi(l+1)-1\right]+\frac{2l+1}{2l}-\ln2,\\
\psi(l)&=\sum_{k=1}^{l-1}\frac{1}{k}-\gamma,
\end{split}
\end{equation}
and $\gamma$ is the Euler constant.
Geodesic orbits in Schwarzschild spacetime can be specified with two orbital elements, the semi-latus rectum $p$ and the eccentricity $e$.
We consider the expansion of quantities in terms of small parameters, $v=\sqrt{1/p}$ and $e$.
For the radial frequency in geodesic equations, the expansion up to $\mathcal{O}(e^2,v^7)$ is
\begin{equation}
\Omega_r=\left(1-\frac{3e^2}{2}\right)v^3+\left(-3+\frac{15e^2}{2}\right)v^5+\left(-\frac{9}{2}+6e^2\right)v^7.
\end{equation}
The total scalar luminosity up to $\mathcal{O}(v^5)$ is,
\begin{equation}\label{aenergy}
\begin{split}
\frac{dE_{\text{scal}}^{\infty}}{dt}=&\left(\frac{dE}{dt}\right)_D\left\{1-2 v^2+2 \pi  v^3-10 v^4+\frac{12 \pi  v^5}{5} \right.\\
&\left.+e^2\left(-1+\frac{158 v^2}{15}+3 \pi  v^3 -\frac{4268 v^4}{105}+\frac{47 \pi  v^5}{3} \right)\right\},
\end{split}
\end{equation}
where
\begin{equation}
\left(\frac{dE}{dt}\right)_D=\frac{d^2}{3}\left(\frac{m_p}{M}\right)^2v^8.
\end{equation}
From Eq.~\eqref{aenergy}, we see that
when the secondary compact object is in the weak field regions $(p>20)$, $v\ll 1$,
the $e^2$ term is negative,
so binaries in more eccentric orbits emit less energy through the scalar field, and the orbital phase difference $\Delta\Phi$ is smaller.
When the secondary compact object is in relative strong field regions $(p<8)$, $v\sim 1$,
the  $e^2$ term becomes positive,
so binaries in more eccentric orbits emit more energy, and the orbital phase difference $\Delta\Phi$ is larger.
If we replace the semi-latus rectum $p$ with the apastron $r_a$, then Eq. \eqref{aenergy} can be rewritten
\begin{equation}\label{aenergy2}
\begin{split}
\frac{dE_{\text{scal}}^{\infty}}{dt}=&\frac{d^2}{3}\left(\frac{m_p}{M}\right)^2\frac{1}{r_a^4}\left\{1-\frac{2}{r_a}+\frac{2\pi}{r_a^{3/2}}-\frac{10}{r_a^2}+\frac{12\pi}{5r_a^{5/2}}\right.\\
&\left.+e\left(4-\frac{10}{r_a}+\frac{11\pi}{r_a^{3/2}}-\frac{60}{r_a^2}+\frac{78\pi}{5r_a^{5/2}}\right)\right\}.
\end{split}
\end{equation}
The $e$ term is always positive, so the orbital phase difference $\Delta \Phi$ is always larger in more eccentric orbits.
The monopolar contribution of the scalar sector is
 \begin{equation}
 \frac{dE_{\text{l=0,m=0}}^{\infty}}{dt}=\left(\frac{dE}{dt}\right)_D \left(\frac{25v^2}{6}-\frac{35  v^4}{2}+      \frac{25}{3} \pi   v^5   \right) e^2.
 \end{equation}
So the monopole radiation only appears for eccentric binaries and the contribution to the luminosity is in the order of $d^2 e^2 v^{10} $.
The dipolar contribution of the scalar sector is
\begin{equation}
\begin{split}
 \frac{dE_{\text{l=1,m=1}}^{\infty}}{dt}=&\left(\frac{dE}{dt}\right)_D \left\{1-\frac{26 v^2}{5}+2 \pi  v^3+\frac{1123 v^4}{175} -\frac{52}{5} \pi  v^5\right. \\
 &\left.+e^2\left(-1+\frac{19 v^2}{10}+3 \pi  v^3+\frac{8496 v^4}{175}-\frac{321}{5} \pi  v^5 \right)\right\}.
 \end{split}
 \end{equation}
In the weak field regions, the dipolar contribution to the luminosity is in the order of $d^2 v^8$ and the eccentricity reduces the contribution.
The quadrupolar contribution of the scalar sector is
\begin{equation}
\begin{split}
 \frac{dE_{\text{l=2,m=2}}^{\infty}}{dt}=&\left(\frac{dE}{dt}\right)_D \left\{\frac{16 v^2}{5}-\frac{848 v^4}{35} +\frac{64 \pi  v^5}{5}\right. \\
 &\left.+e^2\left(\frac{67 v^2}{15}-\frac{4481 v^4}{42}+\frac{1073 \pi  v^5}{15} \right)\right\}.
 \end{split}
 \end{equation}
In the weak field regions, the quadrupolar contribution to the luminosity is in the order of $d^2 v^{10}$ and the eccentricity increases the contribution.
Therefore, the monopolar contribution is negligible and the major contribution is from the dipolar radiation for the scalar sector in the weak field regions.
The gravitational luminosity of the tensor sector is \cite{Tagoshi:1995sh},
\begin{equation}
\begin{split}
\frac{dE_{\text{grav}}^{\infty}}{dt}=&\left(\frac{dE}{dt}\right)_N\left\{1-\frac{1247}{336} v^2+4 \pi  v^3-\frac{44711}{9072} v^4-\frac{8191 \pi }{672} v^5 \right.\\
&\left.+e^2\left(\frac{37}{24}-\frac{65 v^2}{21}+\frac{1087\pi v^3}{48}  -\frac{474409 v^4}{9072}-\frac{118607 \pi  v^5}{1344} \right)\right\},
\end{split}
\end{equation}
where
\begin{equation}
\left(\frac{dE}{dt}\right)_N=\frac{32}{5}\left(\frac{m_p}{M}\right)^2v^{10}.
\end{equation}
In the weak field regions, the total gravitational luminosity is in the order of $v^{10}$,
so the ratio between the scalar and gravitational energy fluxes is in the order of $d^2/v^2$.
As the secondary compact object moves closer to the center massive BH, the ratio becomes smaller.

\section{Parameter estimation}
\label{sec6}
\begin{figure}
\centering
\includegraphics[width=0.9\columnwidth]{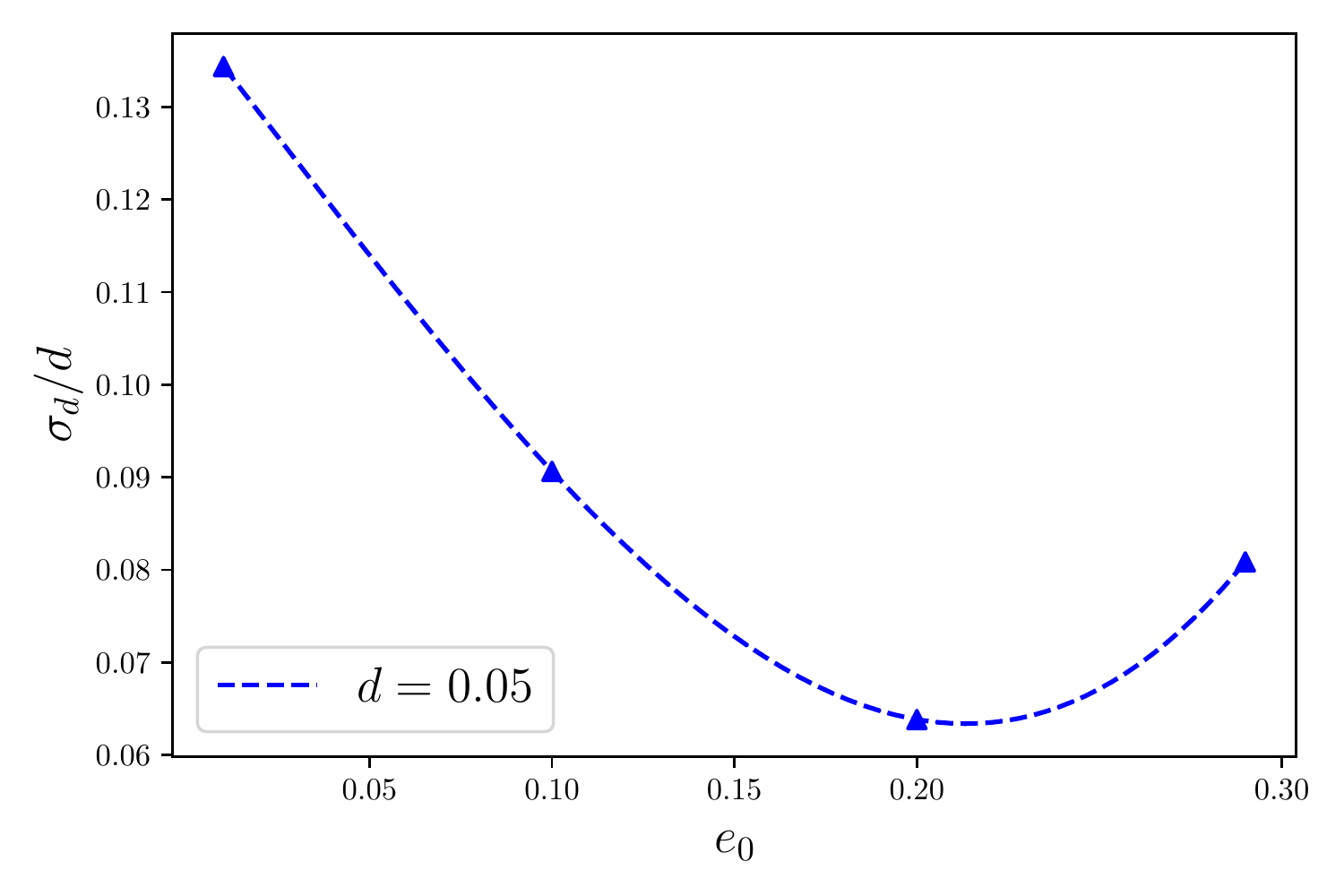}
\caption{
The relative $1-\sigma$ error of the scalar charge $d$ from the one-year observation of EMRIs with $d=0.05$ and different initial eccentricity $e_0$ for LISA.
}
\label{sigmeq}
\end{figure}

\begin{figure}
\centering
\includegraphics[width=0.9\columnwidth]{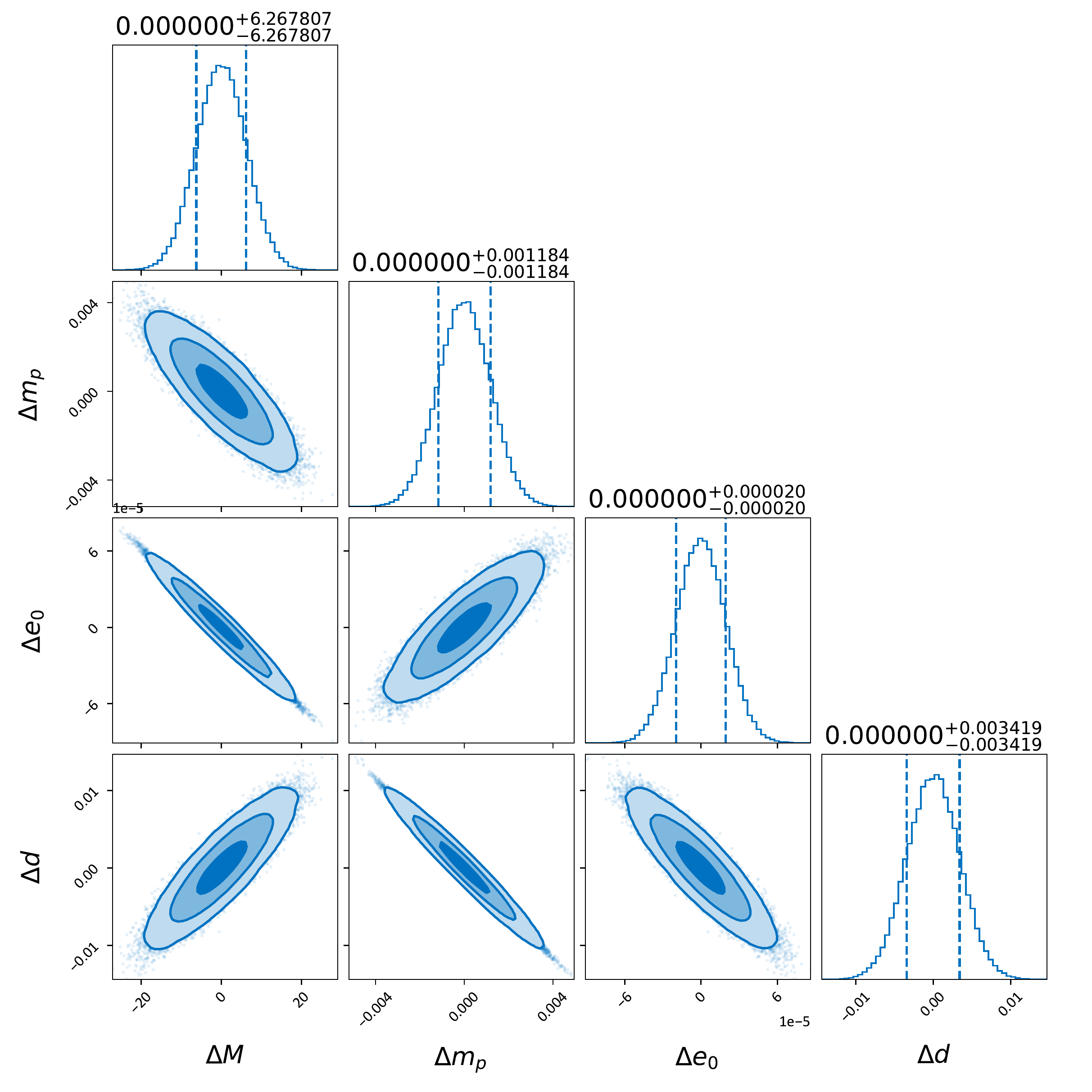}
\caption{The corner plot for the probability distribution of the masses, the semi-latus rectum, eccentricity and electric charge, $(M,m_p, e_0,d)$,
inferred from one-year observation of EMRIs with $d=0.05$ and $e_0=0.2$ for LISA.
Vertical lines show the $1-\sigma$ interval for each source parameter.
The contours correspond to $68\%$, $95\%$ and $99\%$ probability confidence intervals.}
\label{cornere02}
\end{figure}

\begin{figure}
\centering
\includegraphics[width=0.9\columnwidth]{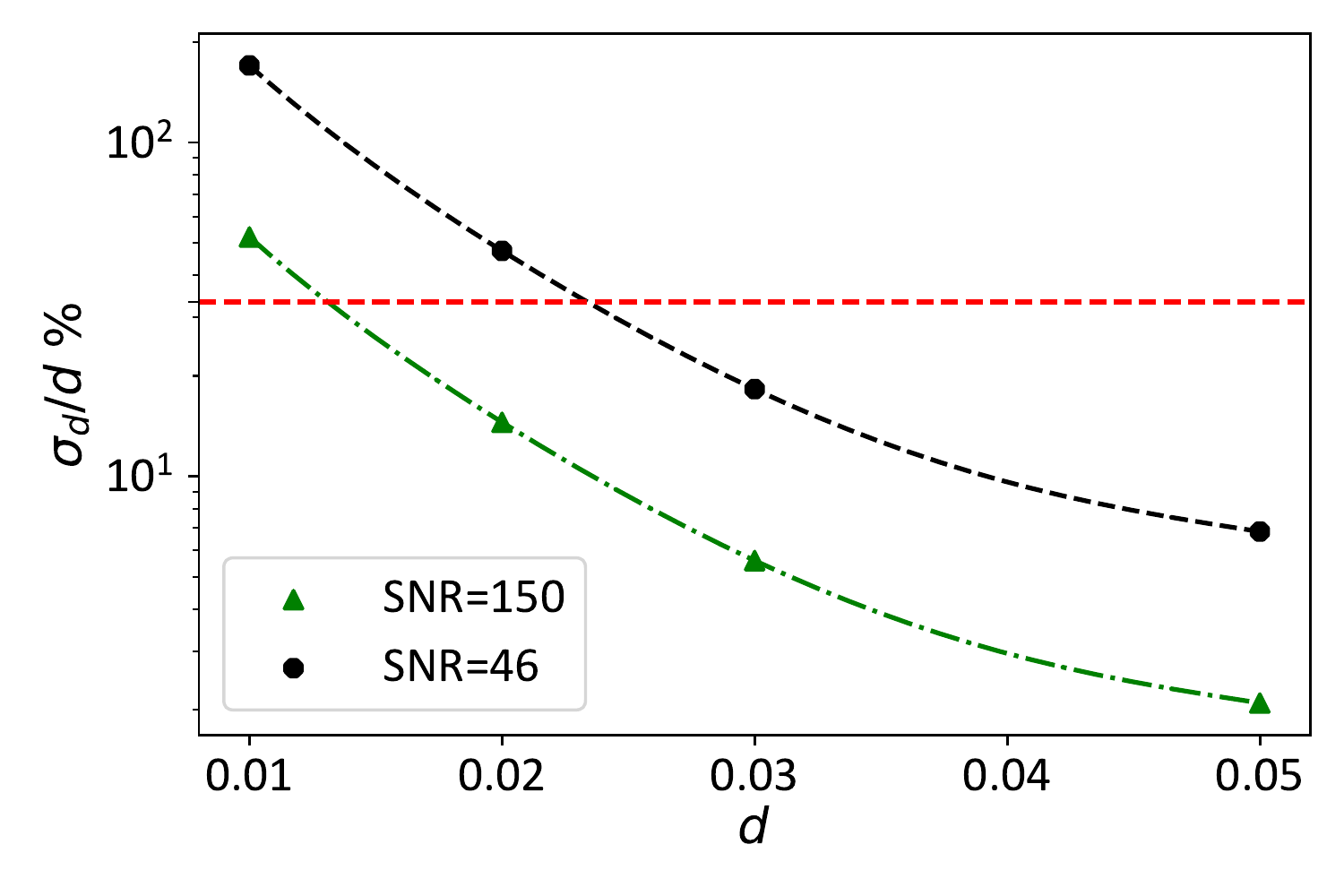}
\caption{
The relative error of the scalar charge $d$ as a function of scalar charge from the one-year observation of EMRIs with eccentricity $e_0=0.2$ for LISA. The horizontal dashed line represents $3-\sigma$ limit and its value is $33.3\%$.
}
\label{sigmed}
\end{figure}

To consider degeneracies among the source parameters and give a more accurate analysis on the detectability of the scalar charge with LISA,
we carry out the parameter estimation with the FIM method.
In the time domain, the GW signal is mainly determined by parameters
\begin{equation}
\xi=(\ln M, \ln m_p, p_0,e_0, d, \theta_s, \phi_s, \theta_1, \phi_1, d_L).
\end{equation}
In the large SNR limit,
the covariances of source parameters $\xi$  are given by the inverse of the Fisher information matrix
\begin{equation}
\Gamma_{i j}=\left\langle\left.\frac{\partial h}{\partial \xi_{i}}\right| \frac{\partial h}{\partial \xi_{j}}\right\rangle_{\xi=\hat{\xi}}.
\end{equation}
The statistical error on $\xi$ and the correlation coefficients between the parameters are provided by the diagonal and non-diagonal parts of ${\bf \Sigma}={\bf \Gamma}^{-1}$, i.e.
\begin{equation}
\sigma_{i}=\Sigma_{i i}^{1 / 2} \quad, \quad c_{\xi_{i} \xi_{j}}=\Sigma_{i j} /\left(\sigma_{\xi_{i}} \sigma_{\xi_{j}}\right).
\end{equation}
Because of the triangle configuration of the space-based GW detector regarded as a network of two L-shaped detectors, with the second interferometer rotated of $60^\circ$ with respect to the first one, the total SNR can be written as the sum of SNRs of two L-shaped detectors \cite{Cutler:1994pb}
\begin{equation}
\rho=\sqrt{\rho_1^2+\rho_2^2}=\sqrt{\left\langle h_1|h_1 \right\rangle+\left\langle h_2|h_2 \right\rangle},
\end{equation}
where $h_1$ and $h_2$ denote the signals detected by two L-shaped detectors.
The total covariance matrix of the source parameters is obtained by inverting the sum of the Fisher matrices $\sigma_{\xi_i}^2=(\Gamma_1+\Gamma_2)^{-1}_{ii}$.
Here we fix the source angles $\theta_s=\pi/3,~\phi_s=\pi/2$, the direction of angular momentum $\theta_1=\pi/4,~\phi_1=\pi/4$, and the initial orbital separation is adjusted to experience one-year adiabatic evolution before the final plunge $r_{\text{end}}=r_{\text{ISCO}}+0.1~M$.
As mentioned above, we consider the EMRI system with $m_p=10~M_{\odot}$, $M=10^6~M_{\odot}$, $d=0.05$, $d_L=1~\text{Gpc}$ and $e_0=(0.01,0.1,0.2,0.29)$.
By adding the scalar flux \eqref{tolflux} into the FastEMRIWaveforms,
we use the SchAAK module to get the signals in LISA.
SNR for LISA is about $\sim46$ with the one-year observation of eccentric EMRIs.
Taking only the intrinsic parameters $\xi=(M,\mu,p_0,e_0,d)$,
we obtain the error $\sigma_d=4.0\times 10^{-3}$ for $e_0=0.29$; $\sigma_d=3.4\times 10^{-3}$ for $e_0=0.2$; $\sigma_d=4.5\times 10^{-3}$ for $e_0=0.1$; and $\sigma_d=6.7\times 10^{-3}$ for $e_0=0.01$.
As shown in Fig. \ref{sigmeq},
the eccentricity helps to reduce the measurement error of the scalar charge and improve the detectability of scalar fields.
Comparing the errors of the scalar charge for $e_0=0.2$ and $e_0=0.29$, 
we find that the error of the scalar charge is larger when the initial eccentricity is $e_0=0.29$.
When $e_0=0.29$, the dephasing between EMRIs with and without scalar charge is smaller, 
and the dephasing becomes the major factor on the detection of the scalar charge compared with the influence on the amplitude shapes of GWs caused by eccentricity, 
thus the estimation error of the scalar charge is larger.
The corner plot of the parameters for $e_0=0.2$ is shown in Fig. \ref{cornere02}.
The results show that the relative measurement error of the scalar charge $d$ is less than 10\%,
$d$ is highly correlated with the mass of the Schwarzschild BH and is anti-correlated with the mass of the small compact object and the initial eccentricity.
The anti-correlation between $d$ and $e_0$ is consistent with the results shown in Fig. \ref{sigmeq}.
The FIM results on the correlation between $d$ and $M$ and the anti-correlation between $d$ and $e_0$ are consistent with the faithfulness analyses shown in Figs. \ref{faithfulness1} and \ref{faithfulness2}.
In order to constrain the scalar charge,
we give the relative error of the scalar charge as a function of the scalar charge with SNR$=46$ and SNR$=150$ in Fig. \ref{sigmed}.
The results show that LISA could detect the scalar charge as small as $d\sim 0.014$ by excluding $d=0$ at the $3\sigma$ confidence level with one-year observation of eccentric EMRIs with SNR=150, and $d\sim 0.023$ with SNR=46.

\section{Conclusions}
\label{sec7}

In modified gravity, BHs may carry scalar charge.
In this paper, we consider the detection of scalar charge carried by the secondary BH in EMRIs as a probe of modified gravity.
We derive the source term $S_{lm\omega}$ of the inhomogeneous Teukolsky equation for the scalar field in eccentric orbits, and calculate the energy and angular momentum emitted by the scalar field.
In order to reduce the numerical error, we subtract out the leading PN behavior from the actual flux values and  interpolate over an effective flux.
We find that scalar fields accelerate the evolution of $p$ due to the additional energy and angular momentum carried away by the scalar fields.
For the orbital parameter $e$, the eccentricity decreases first and then increases near $p_s=6+2e$.
The time that it takes $e$ to reach the minimum is shorter and the values of $p$ and $e$ at the turning point are bigger for larger scalar charge $d$.

Starting from relative strong field regions with smaller orbital distance ($p_0\leq 8$),
the contribution from the eccentricity is positive,
so binaries in more eccentric orbits emit more energy,
and the accumulated dephasing $\Delta\Phi$ is bigger for larger initial eccentricity $e_0$.
However, starting from weak field regions with larger orbital distance ($p\ge 20$),
the contribution from the eccentricity is negative,
so binaries in more eccentric orbits emit less energy through the scalar field,
and the accumulated dephasing $\Delta\Phi$ is smaller for larger initial eccentricity $e_0$.
If instead we use the apastron $r_a$ as the orbital distance,
then the contribution from the eccentricity is always positive,
and the accumulated dephasing $\Delta\Phi$
is always bigger for larger initial eccentricity $e_0$ regardless of the initial values of $r_a$.

We find that the effect of the scalar charge is more significant for eccentric inspirals than circular inspirals
and the mass of the central BH affects the detection of the scalar charge.
For more massive central BHs, for example $M=5\times 10^6~M_{\odot}$,
if the frequency of the higher harmonic is in LISA's sensitive frequency band,
the effect of the eccentricity is likely more significant and improves the detectability of the scalar charge.
By computing the faithfulness between two signals with and without the scalar charge,
we conclude that LISA can detect the scalar charge around $d\ge 0.005$
through one-year observation of EMRIs consisting of a central BH with the mass $M=10^6~M_{\odot}$ and a small BH with the mass $m_p=10~M_{\odot}$.
For EMRIs with $M=5\times10^6~M_{\odot}$ and $m_p=10~M_{\odot}$,
the detection limit for the scalar charge with LISA is $d\ge 0.02$.
By estimating the measurement error of the scalar charge with the FIM method for EMRIs with $M=10^6~M_{\odot}$ and $m_p=10~M_{\odot}$,
we find that LISA could detect the scalar charge as small as $d\sim 0.014$ by excluding $d=0$ at the $3\sigma$ confidence level with one-year observation of eccentric EMRIs with SNR=150 and $d\sim 0.023$ with SNR=46,
$d$ is highly correlated with the mass $M$ of the Schwarzschild BH and is anti-correlated with the mass $m_p$ of the small compact object and the initial eccentricity.
The FIM results confirmed the results obtained with faithfulness,
and tell us that eccentric orbits can help us detect scalar fields.
The correlation between $d$ and $M$ is consistent with the result found in \cite{Maselli:2020zgv} for circular EMRIs with Schwarzschild black hole,
but the correlation between $d$ and $M$ and the anti-correlation between $d$ and $m_p$
are opposite to those found for eccentric EMRIs with the Kerr BH in Ref. \cite{Maselli:2021men}.
This may be due to the effect of the spin of the Kerr BH in the Appendix \ref{appendix} and the reason will be studied in future work.

\begin{acknowledgments}
This work makes use of the Black Hole Perturbation Toolkit package.
The numerical computations were performed at the public computing service platform provided by Network and Computing Center of HUST.
This research is supported in part by the National Key Research and Development Program of China under Grant No. 2020YFC2201504,
the National Natural Science Foundation of
China under Grant Nos. 11875136 and 12147120, and China Postdoctoral Science Foundation
under Grant No. 2021TQ0018.
\end{acknowledgments}

\appendix
\section{Comparison of Fisher information matrix of the Schwarchild case and Kerr case}\label{appendix}
\begin{figure}
\centering
\includegraphics[width=0.9\columnwidth]{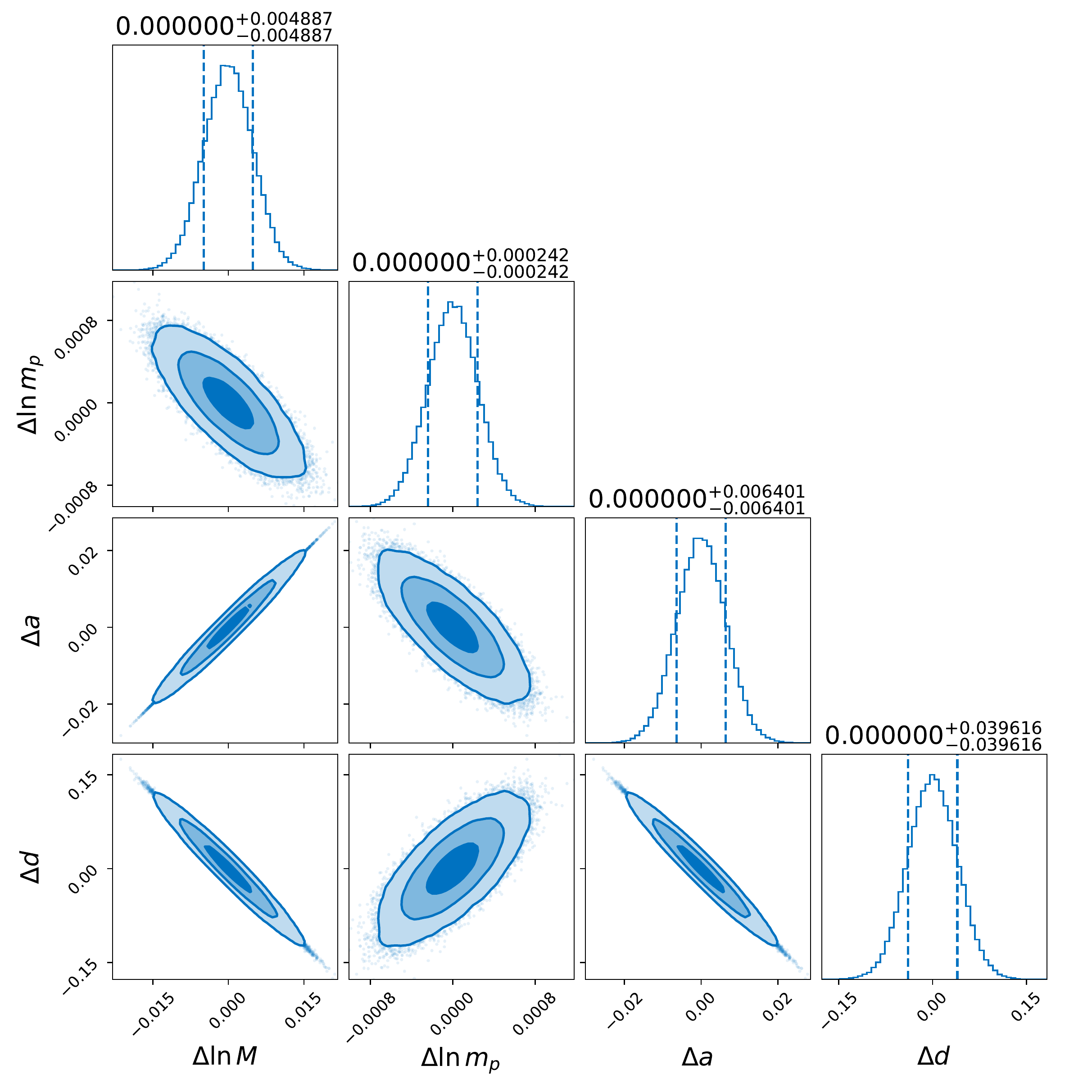}
\caption{Corner plot for the probability distribution of the source parameters $(\ln M,\ln m_p, a, d)$ with LISA, inferred after one-year observations of EMRIs with $d=0.05$ and $a=0$.
Vertical lines show the $1\sigma$ interval for the source parameter.
The contours correspond to the $68\%$, $95\%$, and $99\%$ probability confidence intervals.}
\label{cornera0}
\end{figure}

\begin{figure}
\centering
\includegraphics[width=0.9\columnwidth]{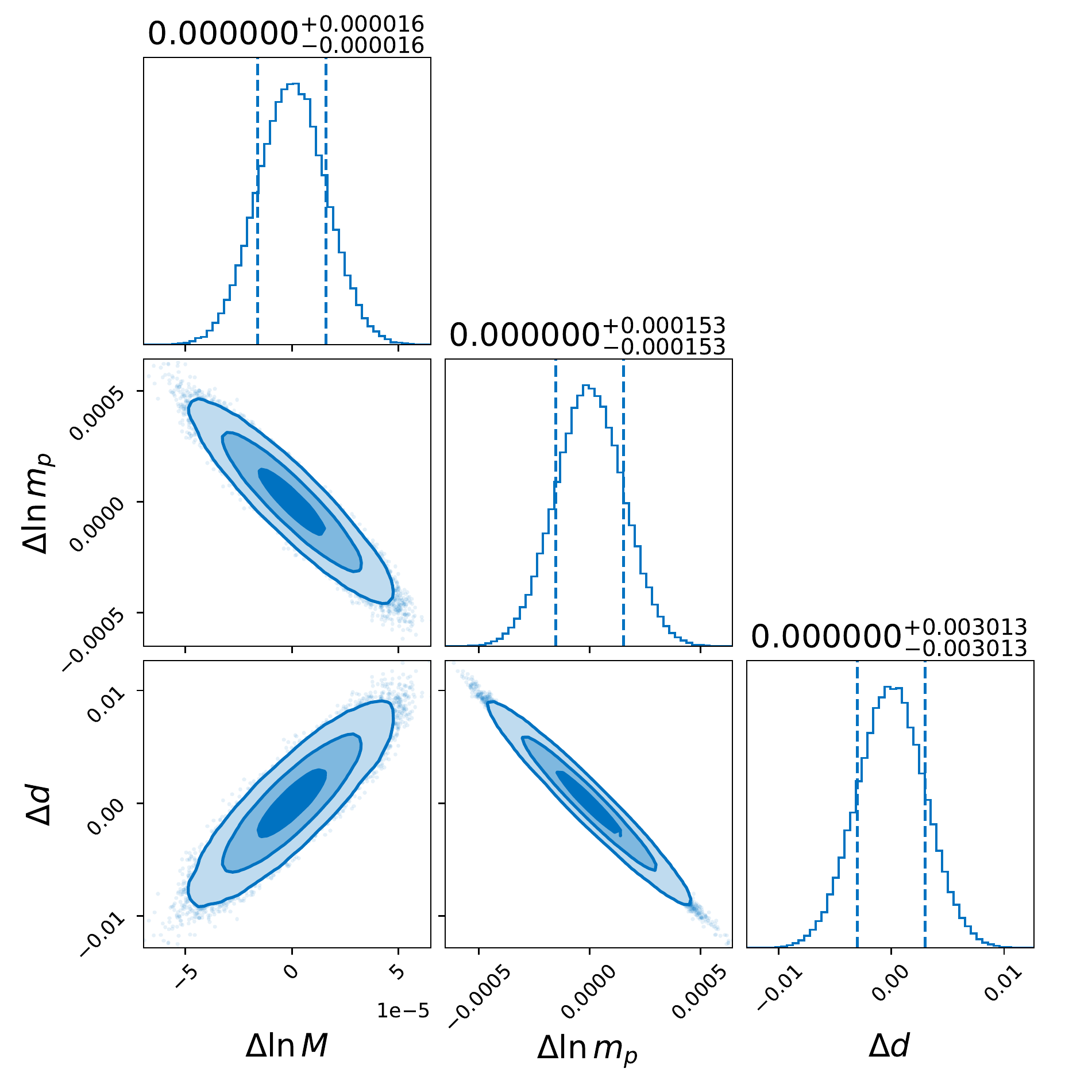}
\caption{Corner plot for the probability distribution of the source parameters $(\ln M,\ln m_p, d)$ with LISA, inferred after one-year observations of EMRIs with $d=0.05$ in the Schwarzschild BH background.
Vertical lines show the $1\sigma$ interval for the source parameter.
The contours correspond to the $68\%$, $95\%$, and $99\%$ probability confidence intervals.}
\label{cornerSCH}
\end{figure}

We consider EMRIs with a scalar charged compact object inspiralling onto a Kerr BH and a Schwarzchild BH in quasi-circular orbits.
Following the procedure \cite{Maselli:2021men}, we give the corner plot for the probability distribution of the masses, primary spin, and secondary charge, $(M, m_p, a, d)$, inferred after one year of observation with LISA for and EMRI with $d=0.05$ and SNR of $150$ in Fig. \ref{cornera0}. 
Off-diagonal panels show that the scalar charge is highly correlated with the secondary mass and anti-correlated with the spin parameter and the mass of the primary, which are consistent with the results in \cite{Maselli:2021men}.
We also compare the corner plot for charged EMRIs in the Schwarzchild BH background in Fig. \ref{cornerSCH}.
The reason is that the spin parameter is associated with parameters such as scalar charge, secondary mass, primary mass, and other parameters. So including the primary spin parameter may influence the relation between $d$ and $m_p$, $d$ and $M$.


\providecommand{\href}[2]{#2}\begingroup\raggedright\endgroup

\end{document}